\documentclass[pra,twocolumn]{revtex4-1}
\usepackage{graphicx}
\usepackage{amsmath}
\usepackage{amssymb}
\usepackage{bm}
\usepackage[T1]{fontenc}
\usepackage{luainputenc}
\usepackage{subfigure}
\usepackage{float}
\usepackage{booktabs}
\usepackage{lipsum}

\newcommand{\bvec}[1]{{\mathbf #1}}

\newcommand{\beq}{\begin{eqnarray}}
\newcommand{\eeq}{\end{eqnarray}}
\usepackage{appendix}
\usepackage{color}

\begin{document}

\title{Machine learning out of equilibrium correlations in the Bose-Hubbard model}
\author{Ali Mokhtari-Jazi and Malcolm P. Kennett}

\affiliation{Department of Physics, Simon Fraser University,\\
Burnaby, British Columbia V5A 1S6, Canada} 

\date{\today}

\begin{abstract}
Calculating the out-of-equilibrium dynamics of many-body quantum systems theoretically is a challenging problem.  Essentially exact results can be obtained for the out-of-equilibrium correlations
in the Bose-Hubbard model in one dimension, but higher dimensions require approximate methods.  One such method is the two-particle irreducible strong coupling (2PISC) approach [M.R.C. Fitzpatrick
and M.P. Kennett, Nucl. Phys. B {\bf 930}, 1 (2018)].  Calculations of the single-particle correlations using this method yield values of the velocity for correlation spreading that match
well with exact methods in one dimension and experiments in one and two dimensions.  However, the 2PISC method is less accurate for determining the amplitude of correlations, especially in the 
regime where interactions are not very strong.  Viewing the calculation of the single-particle correlations as an image correction problem, we train a neural network (NN) to take input from the 
2PISC approach to reproduce the output of exact diagonalization calculations.  We show that the predictions of the NN improve on 2PISC results for parameters outside the training region.  
Our approach is not specific to the Bose-Hubbard model and may find application to the out-of-equilibrium dynamics of other quantum many-body systems.
\end{abstract}

%\begin{abstract} 
%The two-particle irreducible strong coupling (2PISC) approach to the Bose-Hubbard model introduced in Ref. \cite{Fitzpatrick2018a} reproduces the equilibrium and out-of-equilibrium dynamics very well in the strong coupling regime within the Mott phase. However, it loses its accuracy in the weakly interacting regime as the system approaches the Mott insulator-superfluid phase boundary from inside the Mott phase. In the weakly interacting limit, the predictions of the 2PISC approach diverge from the exact diagonalization results, and as the interactions become weaker, the accuracy of the prediction of the amplitude of the single-particle density matrix decreases.
%We use neural network (NN) models to improve the predictions of the 2PISC in the weakly interacting limit. We train the NN with data obtained from the 2PISC and exact diagonalization (ED) for various ranges of the parameter space and use the trained model to predict the single-particle correlations beyond the range of parameter space that is used for the training the NN in order to improve the predictions of the 2PISC in the weakly interacting limit.
%\end{abstract}
\maketitle
\section{Introduction}

The out-of-equilibrium dynamics of quantum many-body systems is a challenging problem that has attracted much recent interest \cite{Polkovnikov2011,Eisert2015}. Ultra-cold atoms in optical lattices
have proven to be a flexible setting to investigate these phenomena \cite{Bloch2005,Lewenstein2007,Bloch2008,Gross2017}.  As a minimal model for interacting bosons in an optical lattice, 
the Bose-Hubbard model (BHM) \cite{Fisher1989,Jaksch} in particular has attracted considerable attention as a setting for studying out-of-equilibrium dynamics \cite{Greiner2002,Bakr2010,Hung2010,Sherson2010,Chen2011,Cheneau2012,Kennett2013,Choi2016,Rubio-Abadal2019,Takasu2020} especially in the context of quantum simulations \cite{Choi2016,Rubio-Abadal2019,Takasu2020}. 

Measurements of the spreading of correlations in the BHM \cite{Cheneau2012,Takasu2020} shed light on how information propagates in this system.  Multiple theoretical methods 
enable the calculation of dynamical correlations in the BHM in one dimension, including exact diagonalization (ED) and time-dependent density matrix renormalization group methods (t-DMRG) \cite{Cheneau2012,Clark2004,Kollath2007,Lauchli2008,Bernier2011,Bernier2012,Barmettler2012,Trotzky2012,Cevolani2018,Despres2019}. However, these tools are not as effective for calculating the 
spreading of correlations in higher dimensions. A variety of methods have been used to study the spreading of correlations in the BHM in two dimensions, 
including Gutzwiller mean-field theory with perturbative corrections \cite{Navez2010,Trefzger2011,Krutitsky2014,Queisser2014}, time-dependent variational Monte 
Carlo \cite{Carleo2014}, doublon-holon pair theories \cite{Yanay2016} and tensor network methods \cite{Kaneko2022}.  An alternative approach to the BHM that we have used is the two particle irreducible (2PI) out-of-equilibrium strong coupling (2PISC) approach \cite{Kennett2011,Fitzpatrick2018a,Fitzpatrick2018b,Fitzpatrick2019,Kennett2020,Mokhtari-Jazi2021}. This approach allows the treatment of the dynamics of the order parameter and correlation functions on an equal footing in dimensions greater than one and we have previously used it to demonstrate excellent agreement \cite{Mokhtari-Jazi2021} with experiments investigating the spreading of correlations for bosons in optical lattices in one and two dimensions \cite{Cheneau2012,Takasu2020}.  It also has the attractive feature that it allows for the inclusion of disorder averaging, which we have used to study the disordered BHM \cite{Mokhtari-Jazi2023}.

The application of machine learning, particularly using neural network (NN) models to study quantum many-body systems has intensified in the past few years \cite{carleo2017solving, carrasquilla2017machine,vanNieuwenburg2017learning, nomura2021helping, melko2019restricted, choo2018symmetries, Hartmann2019, deng2017quantum, torlai2018neural, carleo2019machine,Rem2019,Carrasquilla2021,Carrasquilla2020}.  Neural networks have been used to identify order parameters of different phases of matter \cite{carrasquilla2017machine,vanNieuwenburg2017learning} or to learn the wavefunction of quantum many-body systems \cite{carleo2017solving}.  
There have also been applications to out-of-equilibrium dynamics \cite{Czischek2018,Schmitt2020,Burau2021,Valenti2022,Nelson2022} and learning the phase diagram of the BHM \cite{Saito2017,VargasCalderon2020,HubbardNet2023}.  

We apply a NN method to the out-of-equilibrium dynamics of the BHM. The one-particle density matrix obtained from the 2PISC method is most accurate in the limit of strong interactions, 
but while the phase remains accurate at weaker interaction strengths, the amplitude is much less accurate as interactions weaken approaching the transition from Mott insulator to superfluid.  
We investigate how to improve the accuracy of the calculation of the one-particle density matrix.  We use 2PISC calculations of the one-particle density matrix as an input and train a 
neural network (NN) to reproduce the results of ED calculations and then apply the NN outside of the region of parameter space used for training.  

In particular, we use the U-Net architecture \cite{ronneberger2015unet} which was originally developed for biomedical image segmentation.  The U-Net is able to capture fine details in data with 
precise localization. Our approach is to use the U-net architecture's pattern recognition and data interpolation capabilities to enhance the predictions from the 2PISC approach.  We train 
the U-Net model to learn the discrepancies between the predictions of the 2PISC approach and ED results in the weakly interacting regime.

By training the model on a dataset comprising both the 2PISC predictions and exact results, we enable the U-Net to effectively extrapolate and predict the behaviour of the system for parameter values where the 2PISC model's accuracy diminishes.  
Our approach is not specific to the Bose-Hubbard model and could be 
applied to the out-of-equilibrium dynamics of other quantum many-body systems.

This paper is structured as follows: in Sec.~\ref{sec:model} we introduce the Bose-Hubbard model; in Sec.~\ref{sec:nn} we discuss the details of of the application of the NN model to the BHM; in Sec.~\ref{sec:results} we present numerical results from the NN and in 
Sec.~\ref{sec:disc} we discuss our results and conclude.

\section{The 2PISC approach to the Bose-Hubbard model}
\label{sec:model}

The Hamiltonian for the BHM is
\begin{eqnarray}
\hat{H}_{\rm BHM}&=&-\sum_{\left<i,j\right>}J(t)\left(\hat{a}^\dagger_{\bvec{r}_i} \hat{a}^{\vphantom{\dagger}}_{\bvec{r}_j} + \hat{a}^\dagger_{\bvec{r}_j} 
\hat{a}^{\vphantom{\dagger}}_{\bvec{r}_i}\right) \nonumber\\
&\quad&\quad- \mu \sum_i \hat{n}_{\bvec{r}_i}
+\frac{U}{2}\sum_i \hat{n}_{\bvec{r}_i}\left(\hat{n}_{\bvec{r}_i} - 1\right),
\end{eqnarray}
where $\hat{a}_{\bvec{r}_i}^\dagger$ and $\hat{a}^{\vphantom{\dagger}}_{\bvec{r}_i}$ are the bosonic creation and annihilation operators for a boson located on site $i$ at location $\bvec{r}_i$, 
$\hat{n}_{\bvec{r}_i}$ is the number operator, $U$ is the interaction strength and $\mu$ is the chemical potential. We allow for time dependent hopping $J(t)$ restricted to nearest
neighbour sites and take the lattice spacing to be $a$. 
The effective theory (ET) and equations of motion obtained with the 2PISC approach were presented in Refs.~\cite{Fitzpatrick2018a,Fitzpatrick2018b,Mokhtari-Jazi2021}.
One can solve the equations of motion that arise from the effective theory to obtain the single particle density matrix
\begin{eqnarray}
	\rho_1(\Delta\bvec{r} = \bvec{r}_i - \bvec{r}_j,t) & = & \left<\hat{a}^\dagger_{\bvec{r}_i}(t) \hat{a}^{\vphantom{\dagger}}_{\bvec{r}_j}(t)\right>,
\end{eqnarray}
 which contains all the information about single-particle observables,
and on a lattice can be written in the form
\begin{eqnarray}
 \rho_1(\Delta\bvec{r},t)  =  \frac{1}{N_s} \sum_{\bvec{k}} \cos\left(\bvec{k}\cdot\Delta\bvec{r}\right) n_{\bvec{k}}(t),
\end{eqnarray}
where $n_{\bvec{k}}(t)$ is the particle distribution over the quasi-momentum $\bvec{k}$ at time $t$, which is related to the density 
\begin{eqnarray}
        n(t) & = & \frac{1}{N_s} \sum_{\bvec{k}} n_{\bvec{k}}(t),
\end{eqnarray}
where $N_s$ is the number of sites.

In the limit of small $J/U$ the ground state is a Mott insulator, while for large $J/U$ the ground state is a superfluid, 
and there is a quantum phase transition between the two.  To study the spreading of correlations in this model, a standard protocol is to start in the Mott insulator at 
unit filling in the limit $J/U \sim 0$ and then ramp the hopping to $J_f$ over a short timescale.  Experimentally, $J_f$ is usually chosen so that the corresponding ground state is Mott insulating \cite{Cheneau2012,Takasu2020}.  The quench protocol we follow is to start with $J/U$ = 0 for a $\bar{n}$ = 1 Mott phase at time $t=0$ and then ramp $J$ to a final value $J_f$ over a
timescale $\tau_Q$ , with the time $t_c$ marking the midpoint of the
quench \cite{Fitzpatrick2018b}. We solve the ET equations of motion to obtain
$\rho_1(\Delta \bvec{r}, t)$.

\begin{figure}[htb]
\subfigure{
    \includegraphics[width=0.4\textwidth]{ 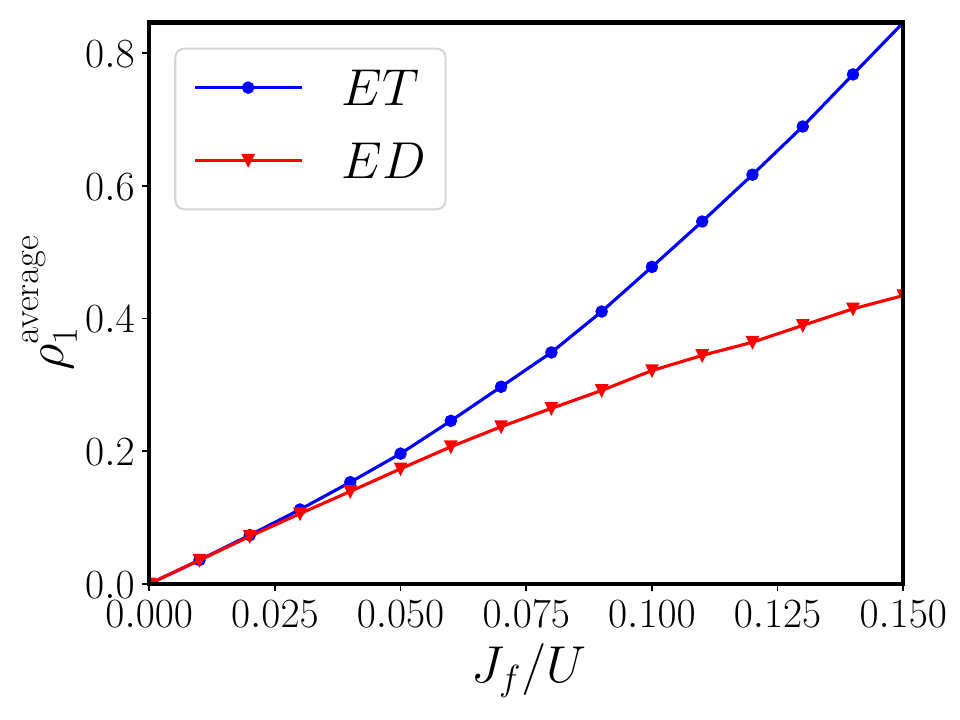}
    }
    \caption{Comparison of exact diagonalization (ED) calculations and effective theory (ET) calculations of the time-averaged $\rho_1(\Delta {\bvec{r}}=a,t)$ up to time $t/U^{-1}=50$.
         The parameters are $L=10$, $\beta U = 1000$, $\mu/U = 0.4116$, $t_c /U^{-1} = 5$, and $t_Q/ U^{-1} = 0.1$.}
    \label{fig: ED ET comparing}
\end{figure}

Truncations in the ET mean that it is not exact, and one consequence is that while one would usually expect the total 
particle number to be conserved, it has small fluctuations.  These do not appear
to affect the determination of the velocity at which correlations spread using the single particle density matrix \cite{Fitzpatrick2018b}, which match extremely 
well with exact results in one dimension \cite{Barmettler2012}.  However, as noted in Ref.~\cite{Mokhtari-Jazi2021}, the amplitude of the single particle density matrix, 
while in good agreement with exact diagonalization results for small $J_f/U$ is in much less good agreement for larger values of $J_f/U$.  A comparison between the time averaged
value of $\rho_1(|\Delta\bvec{r}| = a,t)$ calculated using the ET and ED is shown in Fig.~\ref{fig: ED ET comparing}.

\section{Neural Network for out-of-equilibrium dynamics of the Bose-Hubbard model}
\label{sec:nn}
The effective theory discussed in Ref.~\cite{Fitzpatrick2018a} allows us to calculate $\rho_1(\Delta {\bvec{r}},t)$ throughout the Mott phase and for much larger size systems than are available
to exact diagonalization calculations. However, as illustrated in Fig.~\ref{fig: ED ET comparing}, as $J_f/U$ increases, the amplitude of $\rho_1$ calculated from the 
ET becomes increasingly inaccurate.  This motivates the work here in which we train a NN with ET calculations as the input to obtain improved estimates of the single particle density matrix.

To improve the agreement between the predictions of the 2PISC approach and ED, we use a U-Net NN with a tailored architecture. 
Our approach is to train the NN to predict accurately the $\rho_1(\Delta\bvec{r},t)$ that one would obtain from ED given the approximate $\rho_1(\Delta\bvec{r},t)$ from the ET as input.
Essentially, our goal is to train the NN to predict the correction terms to our ET calculation, while maintaining a sufficiently low generalization error such that the NN can make accurate predictions
of these correction terms in regions of parameter space where the ET is less accurate.

The U-Net architecture utilizes an encoding-decoding structure to effectively correlate and reconstruct the input data relative to the output. In the encoding phase, the single-particle density matrix obtained from the ET is processed through successive layers of convolution, which identify and enhance important features using various filters. This is followed by pooling, which reduces the dimensionality of the data, simplifying the information while preserving the most important image features. These steps compress the data and highlight the critical features that differentiate the input from the desired output. This process aids in detecting patterns that reveal discrepancies between the 2PISC predictions and ED results.
In the decoding phase, U-Net employs transposed convolutions to expand the feature maps, leading to enhanced resolution of the output. 
The integration of skip connections from the encoding layers to the decoding layers ensures that both spatial information and high-level semantic features are preserved and utilized.

%Without the initial data from the effective theory, it would be challenging for the NN to make extrapolations based only on ED results, hence using results from the effective theory as the starting point significantly improves the NN's predictive accuracy.
  
The quantitative agreement between the ET and ED is excellent at small values of $J_f/U$, but becomes less accurate for values of $J_f/U$ close
to the superfluid-Mott insulator transition, where there is a discrepancy in the magnitude of $\rho_1(\Delta {\bvec{r}},t)$ by roughly a factor of 2.  However,
the phase of  $\rho_1(\Delta {\bvec{r}},t)$, in particular, the position of the first peak, is represented accurately by the 
ET \cite{Mokhtari-Jazi2021}. In Fig. \ref{fig: ED ET comparing}, we compare average values of the magnitude of the single-particle correlations 
obtained from the two different methods in one dimension for a chain lattice of length 10 as a function of $J_f/U$ illustrating that the discrepancy grows 
with increasing hopping amplitude. 

The next subsections explain the method, input, output, architecture, and parameters for the U-Net model in more detail.

\subsection{Method}

\begin{figure}[t]
\includegraphics[width=0.4\textwidth]{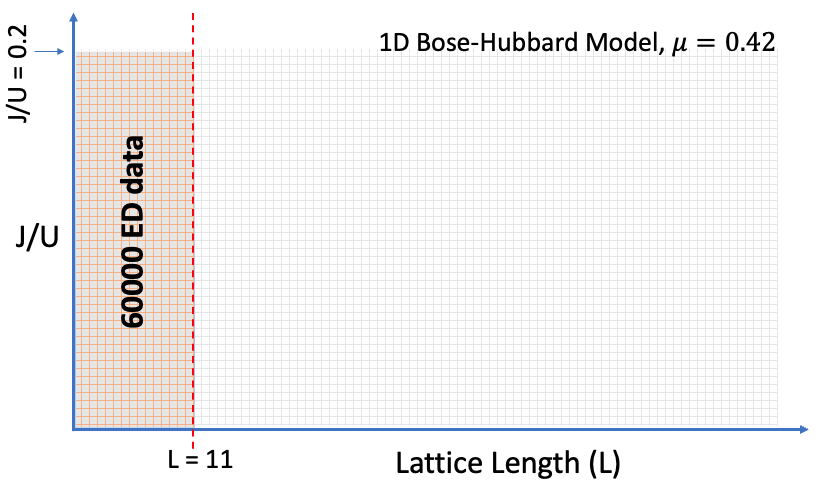}
    \caption{Illustration of the subset of parameter space in which we obtain exact diagonalization data. $\mu/U= 0.42$, $t_c/U^{-1}$ and $\tau_Q/U^{-1}$}
    \label{fig: ED parameter space}
\end{figure}

We utilize supervised learning for a specific subset of parameter space where exact diagonalization is feasible. For this purpose, we generate \(60,000\) instances of \(\rho_1(\Delta \mathbf{r},t)\) across various values of lattice length \(L\), displacement vector \(\Delta\mathbf{r}\), and hopping amplitude \(J\). 
ED is limited by system size \(L\) and we consider all integer values ranging from \(2\) to \(11\). Due to the implementation of periodic boundary conditions in our lattices, the values of \(\Delta\mathbf{r}\) are constrained to integers from \(1\) to \(\left\lfloor\frac{L}{2}\right\rfloor\). As for the hopping amplitude \(J\), we explore values from \(0.0001\) to \(0.2\) in increments of \(0.0001\). The combination of these three parameters yields a dataset of \(60,000\) samples. For all samples, certain parameters remain constant: \(\mu/U = 0.4116\), \(t_c /U^{-1} = 5\) and \(t_Q/ U^{-1} = 0.1\). Figure \ref{fig: ED parameter space} shows the region of parameter space where we obtain our ED data in the $L$-$J_f/U$ plane.

In our supervised learning approach, we train our neural network using 60,000 samples of $\rho_1(\Delta\mathbf{r},t)$ generated from the ET with the same parameters as used for ED. These samples serve as the input to the NN, while the ED results act as the target output. The NN's task is to match each input (ET approximation) with the target (ED result) and learn the necessary adjustments to align the inputs with the correct outputs.

One aspect of our method is that the NN operates without specific knowledge of system parameters like $L$, $\Delta\mathbf{r}$, or $J_f/U$. In the strongly interacting regime, the ET predictions are very close to the ED results and require minimal adjustment. As the value of $J_f/U$ increases, indicating weaker interactions, the predictions need more substantial corrections. The NN 
learns to apply the right amount of modification to the input, without knowing the actual $J_f/U$ value, effectively discerning when and how much to adjust.

\subsection{Architecture and parameters}
We developed a tailored U-Net architecture for our problem. The input is an $n$-dimensional array, and the output is an array with the same dimensionality. In the 
original implementation of the  U-Net \cite{ronneberger2015unet}, the input was a set of images represented as 2-dimensional arrays. We take the inputs and outputs of our implementation to be 
one-dimensional arrays of size $800$ corresponding to $\rho_1(\Delta {\bvec{r}},t)$ for $0<t/U^{-1}<40$. 

\begin{figure}[ht]
  \includegraphics[width=8cm]{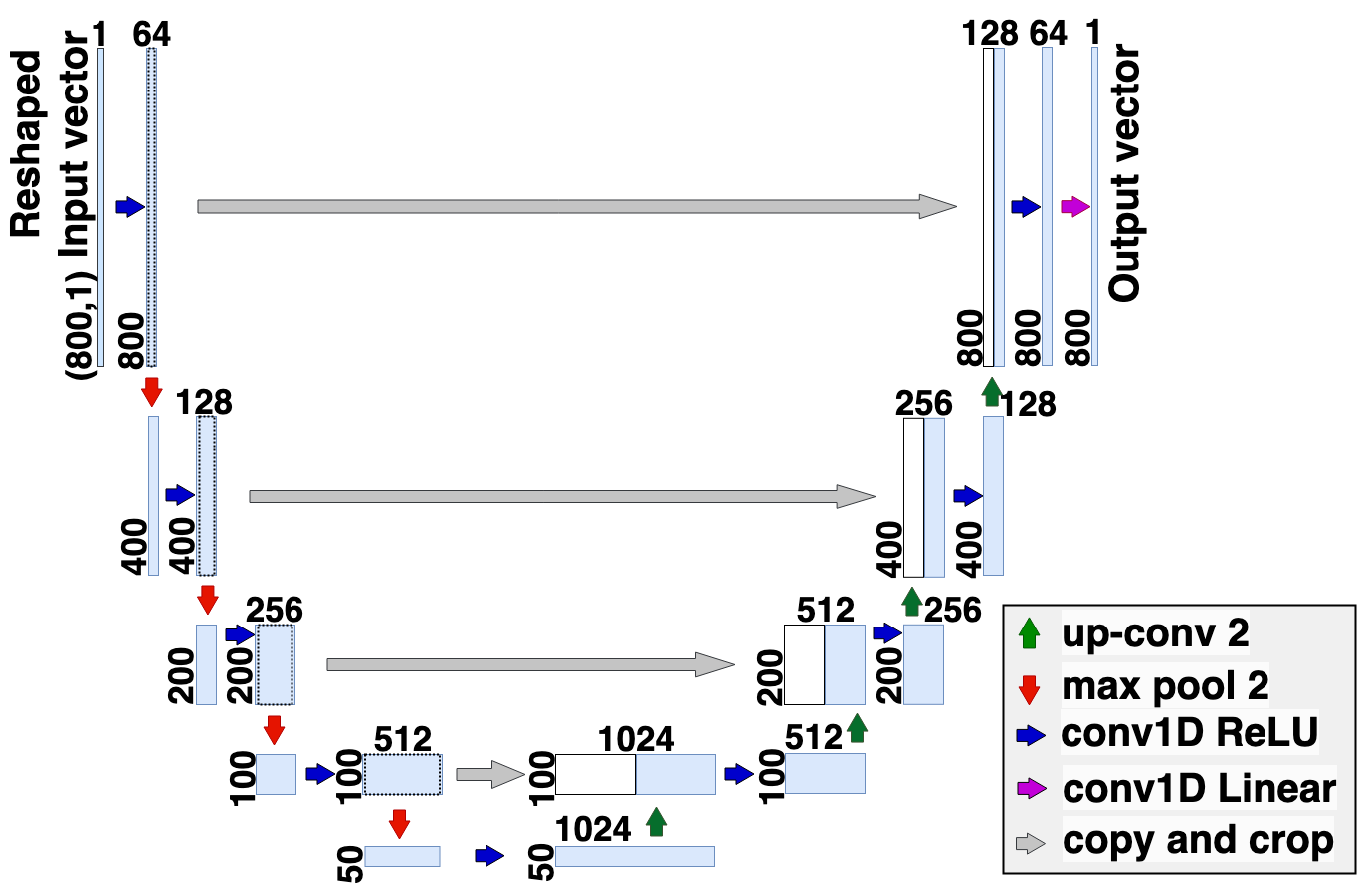}
  \caption{Schematic of the U-Net neural network architecture implemented here.}
  \label{fig:unet}
\end{figure}

We designed our U-Net model for one-dimensional signal processing, utilizing a sequential arrangement of convolutional and pooling layers as illustrated in Fig.~\ref{fig:unet}.
The network initiates with a convolutional layer that contains 64 kernels of size 35, with the ReLU activation function to extract features from the input signal. This layer is immediately followed by a max pooling layer, which downscales the signal by half, thus retaining only the most significant features and introducing translational invariance.

As the signal progresses through the network, each subsequent convolutional layer doubles the number of filters from the previous one, running from 64 to 1024. 
Each convolutional layer is paired with a max pooling layer to continue the pattern of downscaling and feature abstraction.

After converging to the most compressed representation, the network architecture then shifts to an expansive path. Here, we employ transposed convolutional layers to incrementally increase the size of the feature maps. With each upsampling operation, the number of filters is reduced by half, descending from 1024 back to 64. These layers are merged with equivalent feature maps from the contracting path via concatenation, restoring spatial resolution and detail lost in the downsampling stages. We use a stride of 2 for the transposed convolutional layers to ensure proper scaling.
The kernel size is 35 for each layer except the final one, where the output is generated by a convolutional layer with a single kernel of size 1, applying a linear activation function to produce the processed signal. This U-Net architecture effectively leverages the strengths of convolutional layers for feature extraction and transposed convolutions for spatial reconstruction, making it well-suited for detailed signal analysis. 

The model has \(73,119,809\) trainable weight parameters in total. We used a batch size of 16. Training began with an initial learning rate of \(0.0001\), which was adaptively decreased in response to the reduction in training error, see Appendix \ref{app:training_vs_validation_loss}. For optimization, the Adam algorithm was utilized
%, noted for its computational efficiency and adeptness at handling objectives that change over time. The 
and the chosen cost function was 
%\(\textit{LogCosh()}\), which is the logarithm of the hyperbolic cosine of the prediction error as follow
\begin{equation}
    \text{logcosh}(y, \hat{y}) = \sum_{i=1}^{m} \log\left[\cosh(y_i - \hat{y}_i)\right],
    \label{eq: logcosh loss}
\end{equation}
where $y$ and $\hat{y}$ are are the actual and predicted values, respectively.

\section{Neural Network results}
\label{sec:results}
After training our NN, we tested its ability to extrapolate to parameter values from outside the training region.  We first present
results in one dimension for the parameter region depicted in Fig. \ref{fig: ED parameter space}, and then show results in two dimensions.

We maintained a learning rate of 0.0001, which was reduced by half if performance did not improve after one epoch, using the "ReduceLROnPlateau" scheduler. Training consisted of 8 epochs with a batch size of 16. The Adam optimizer was employed for optimization. The test set to validation set ratio was 0.05.

\begin{widetext}

\begin{figure}[bht]
\subfigure{
\includegraphics[width=0.3\textwidth]{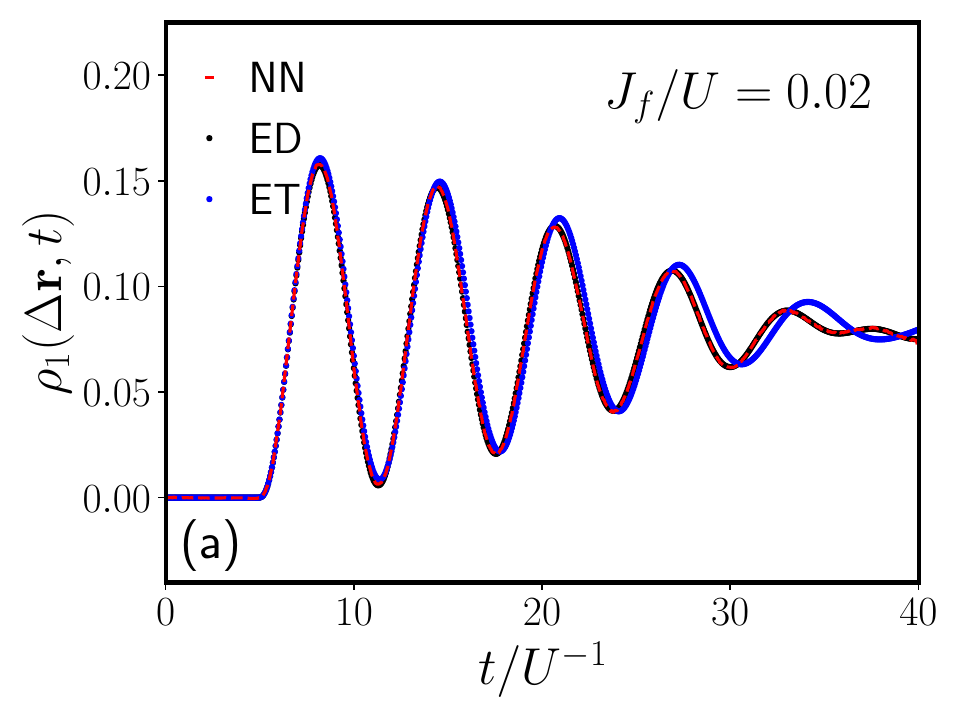}
}
\subfigure{
\includegraphics[width=0.3\textwidth]{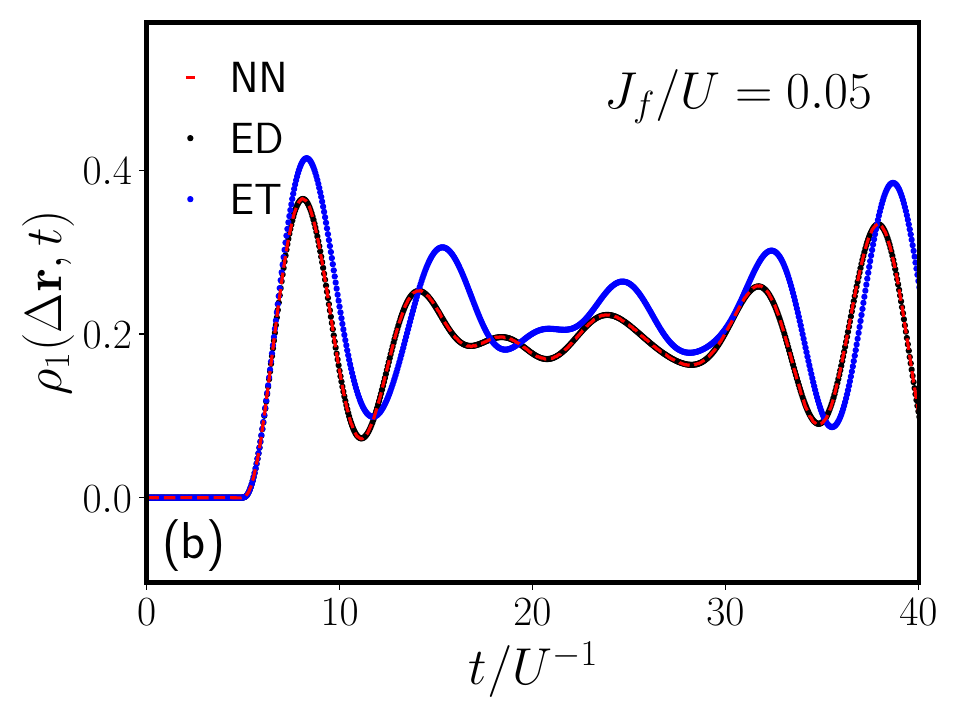}
}
\subfigure{
\includegraphics[width=0.3\textwidth]{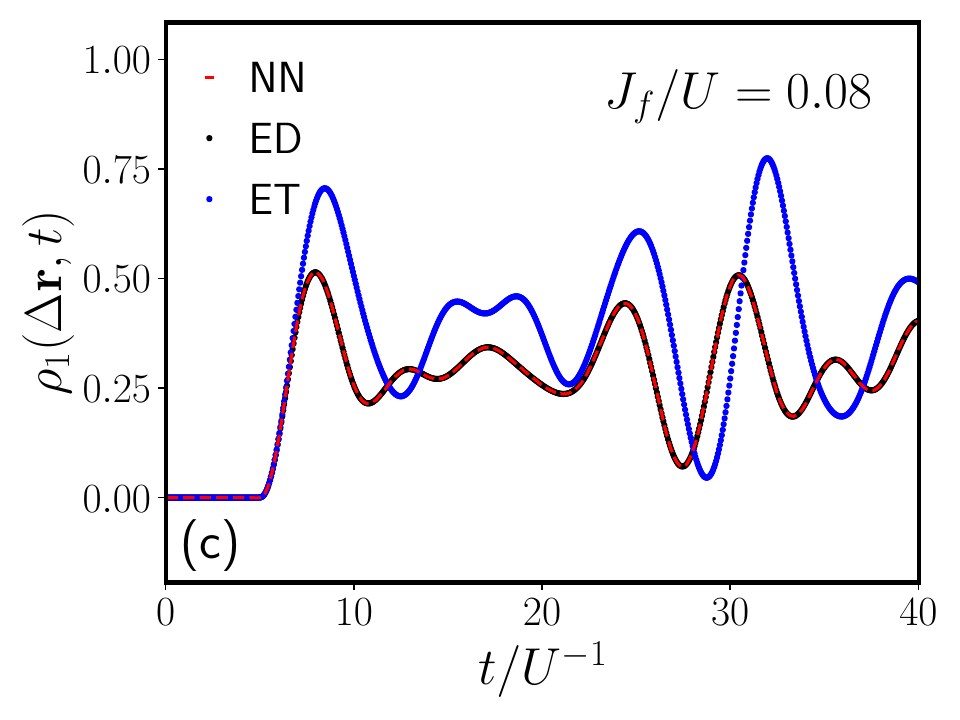}
}
\caption{Comparison of the single-particle density matrix $\rho_1\left(\Delta\mathbf{r}, t\right)$ obtained from the effective theory (ET), exact diagonalization (ED) and the neural network (NN) model for a one-dimensional lattice with $L=10$ and $\Delta\mathbf{r}=1$. Sub-figures (a) to (c) correspond to hopping amplitudes $J_f/U$ with magnitudes of 0.02, 0.05, and 0.08, respectively. The remaining parameters used were $\mu/U = 0.42$, $t_c /U^{-1} = 5$, and $t_Q/ U^{-1} = 0.1$. }
\label{fig: comparing ED ET NN interapolation}
\end{figure}

\end{widetext}

\begin{widetext}

\begin{figure}[thb]
\subfigure{
\includegraphics[width=0.3\textwidth]{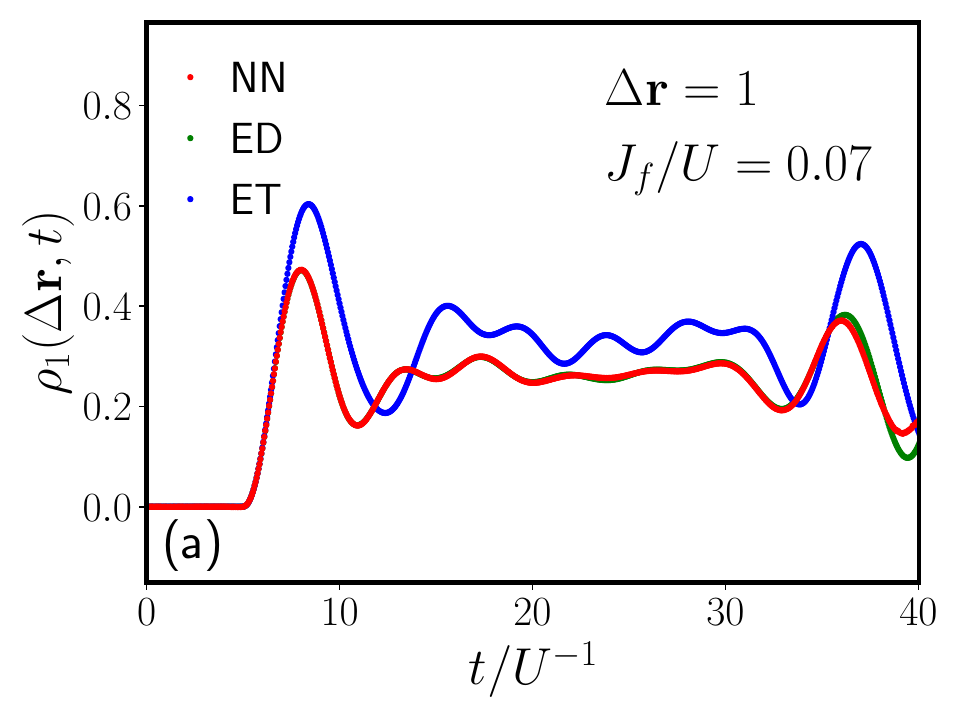}
}
\subfigure{
\includegraphics[width=0.3\textwidth]{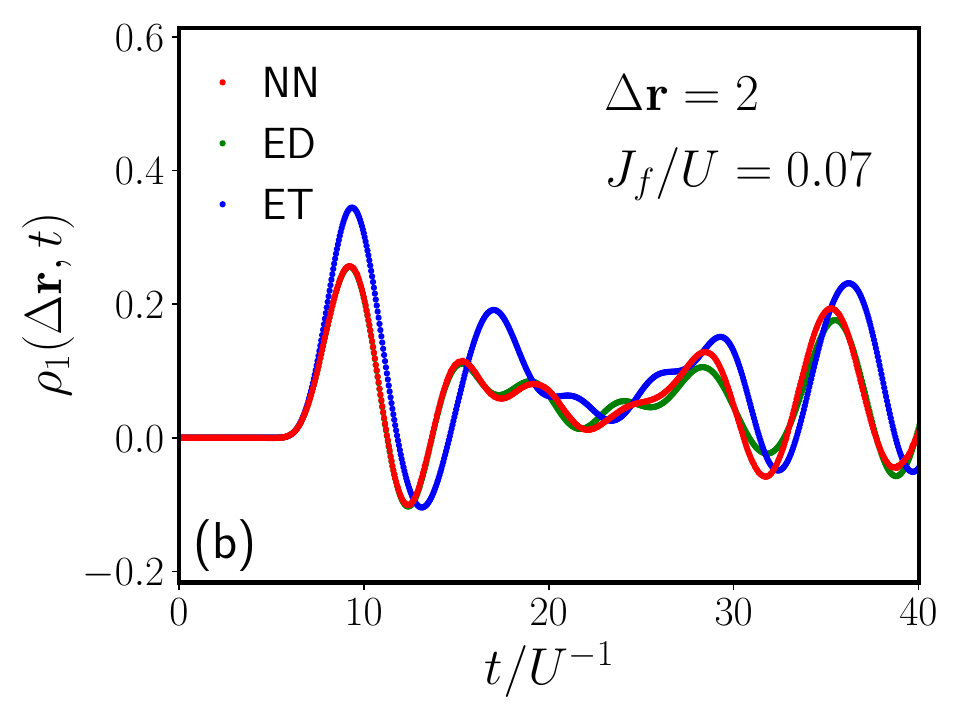}
}
\subfigure{
\includegraphics[width=0.3\textwidth]{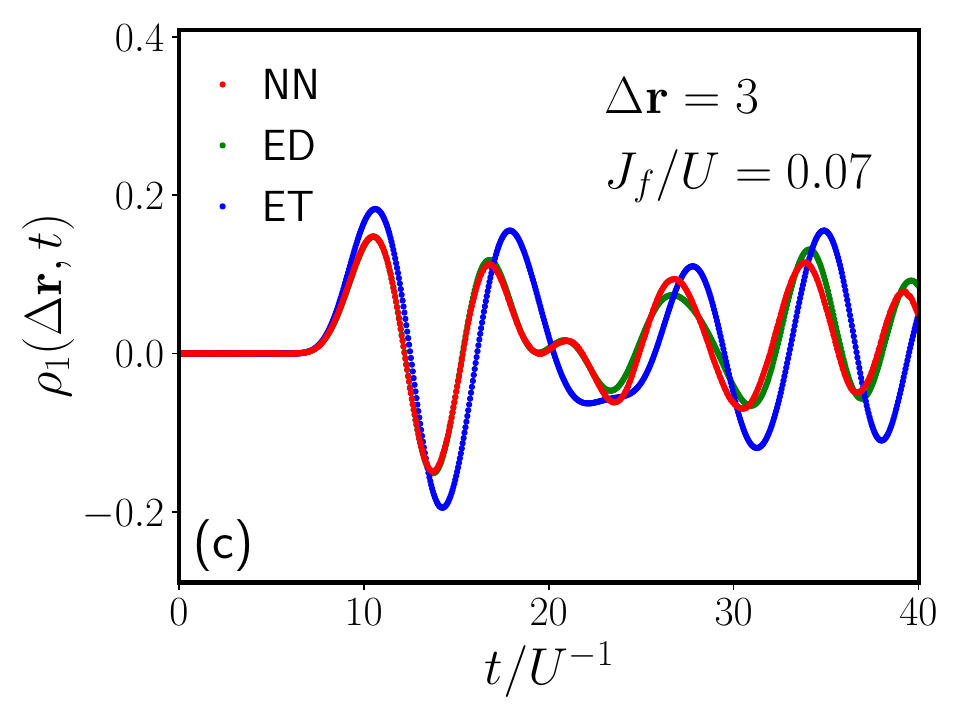}
}
\subfigure{
\includegraphics[width=0.3\textwidth]{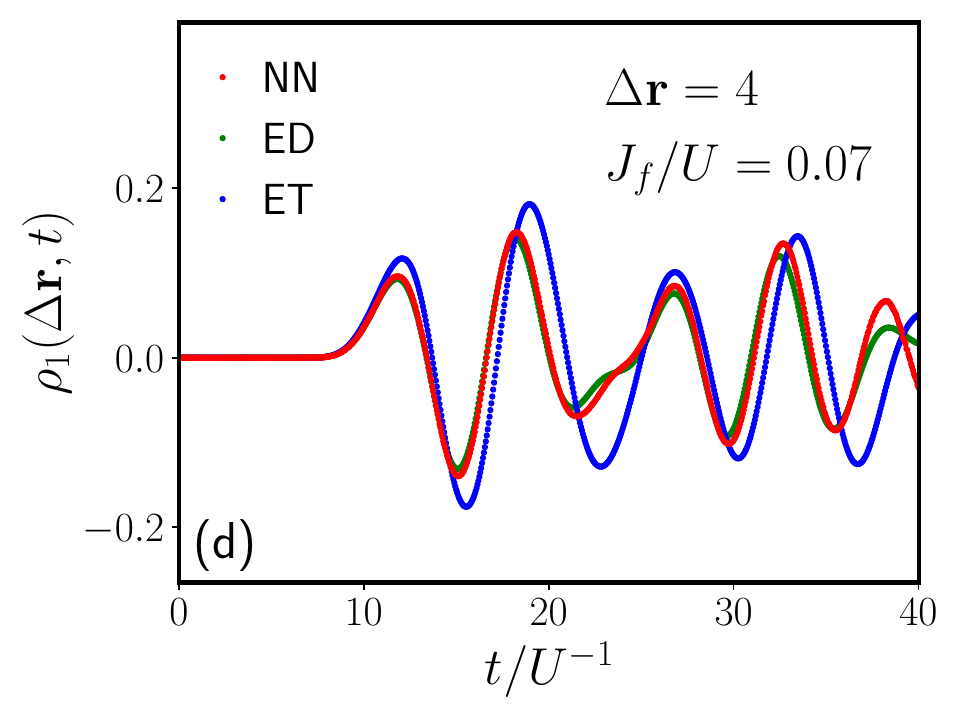}
}
\subfigure{
\includegraphics[width=0.3\textwidth]{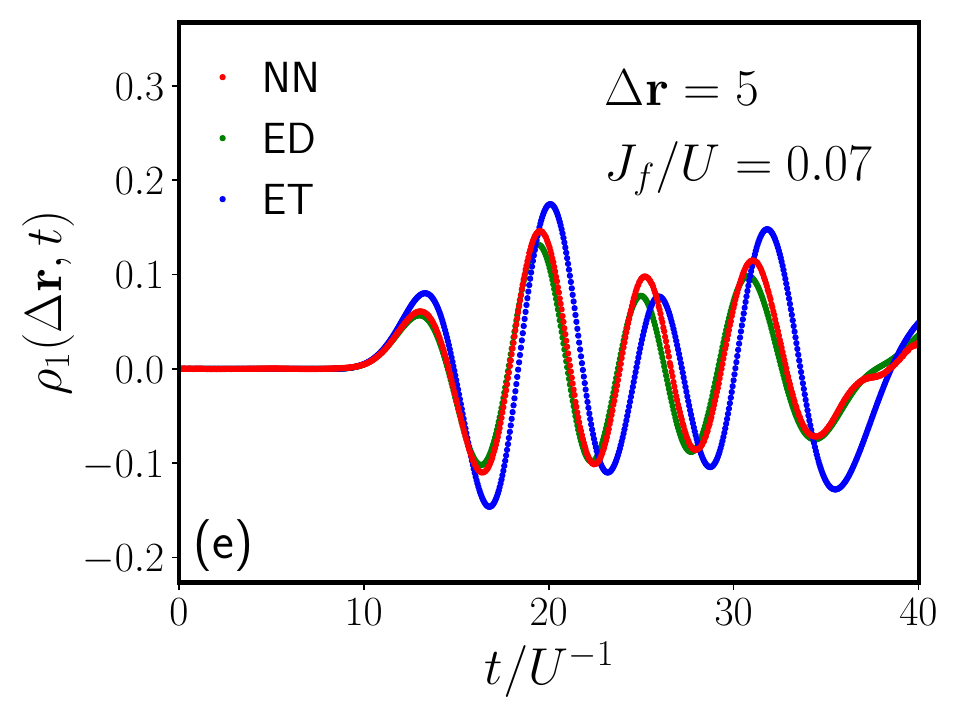}
}
\subfigure{
\includegraphics[width=0.3\textwidth]{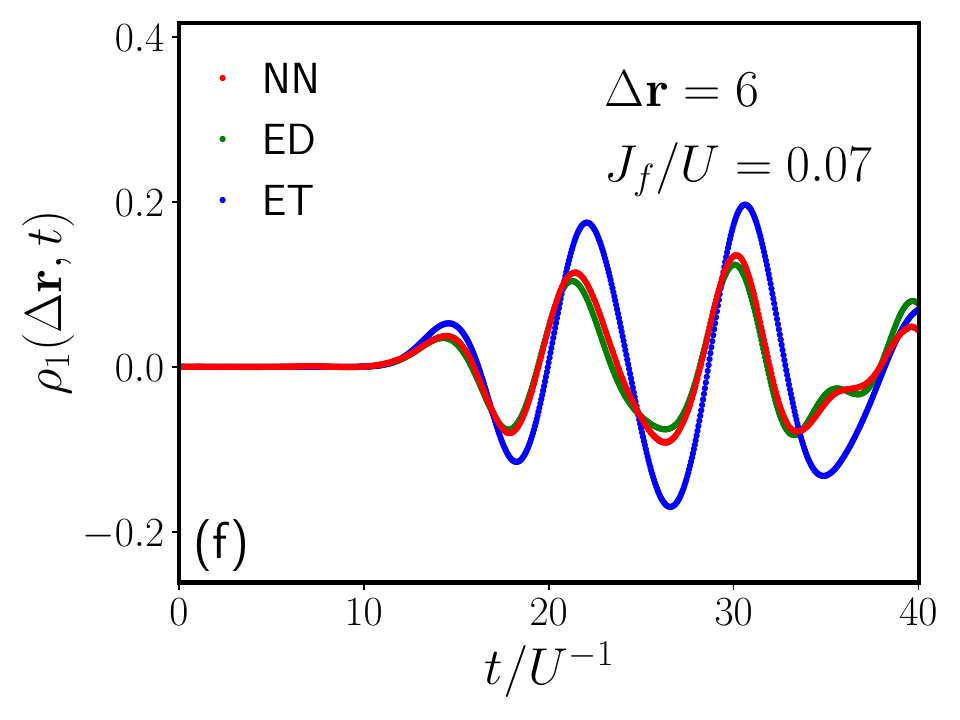}
}
\caption[Comparing NN, ED, and ET for J=0.07 and L=14]{Comparison of the single-particle density matrix $\rho_1\left(\Delta\mathbf{r}, t\right)$ obtained from the effective theory (ET), exact diagonalization (ED) and the neural network (NN) model for a one-dimensional lattice with $L = 14$. Sub-figures (a) to (f) correspond to displacement vectors $\Delta\mathbf{r}$ with magnitudes of 1 to 6, respectively. The remaining parameters used were $\mu/U = 0.42$, $J_f/U = 0.07$, $t_c /U^{-1} = 5$, and $t_Q/ U^{-1} = 0.1$. }
\label{fig: NN 14 J=0.07}
\end{figure}
    
\begin{figure}[htb]
\subfigure{
\includegraphics[width=0.3\textwidth]{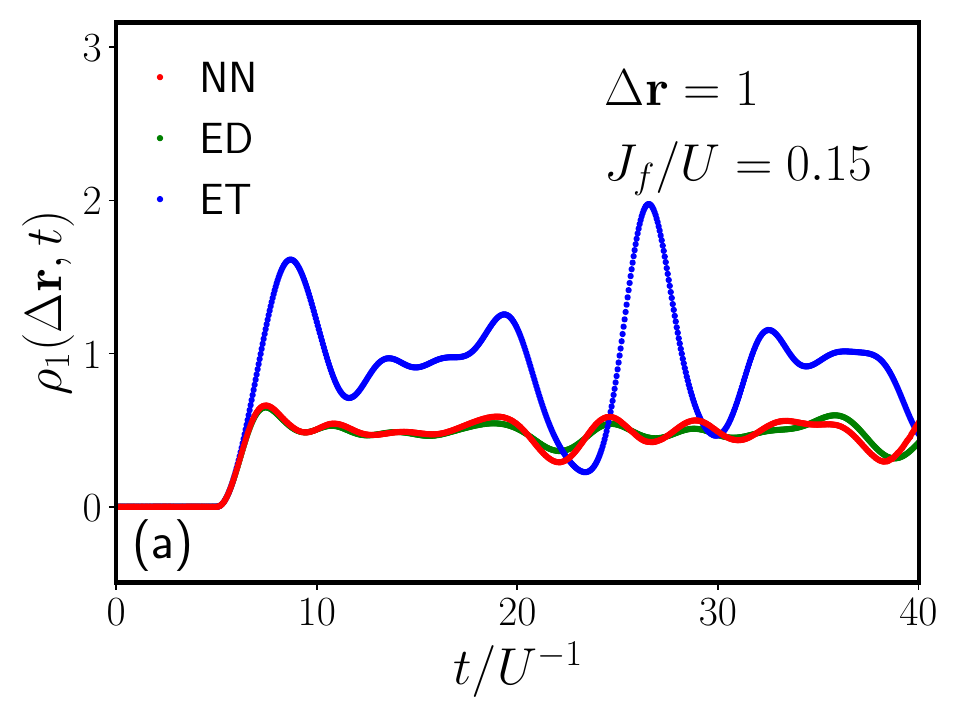}
}
\subfigure{
\includegraphics[width=0.3\textwidth]{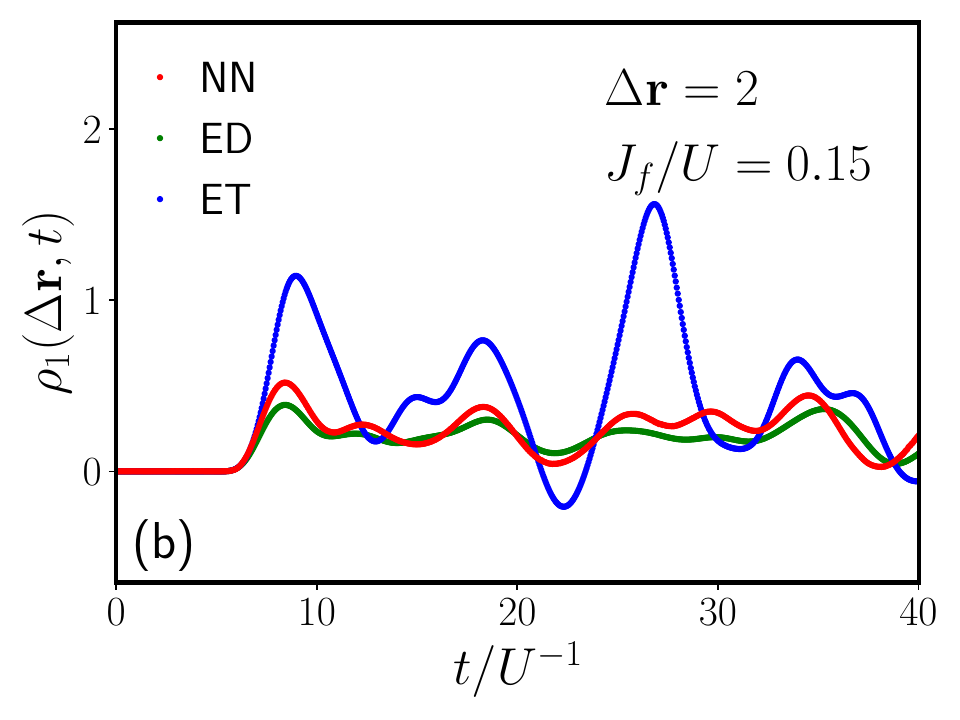}
}
\subfigure{
\includegraphics[width=0.3\textwidth]{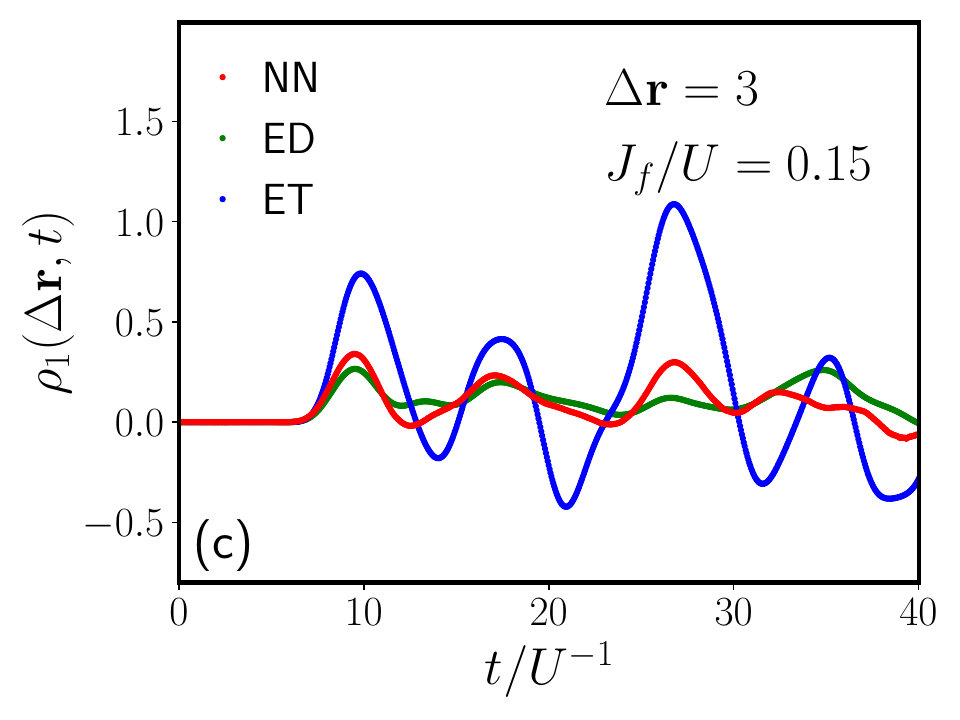}
}
\subfigure{
\includegraphics[width=0.3\textwidth]{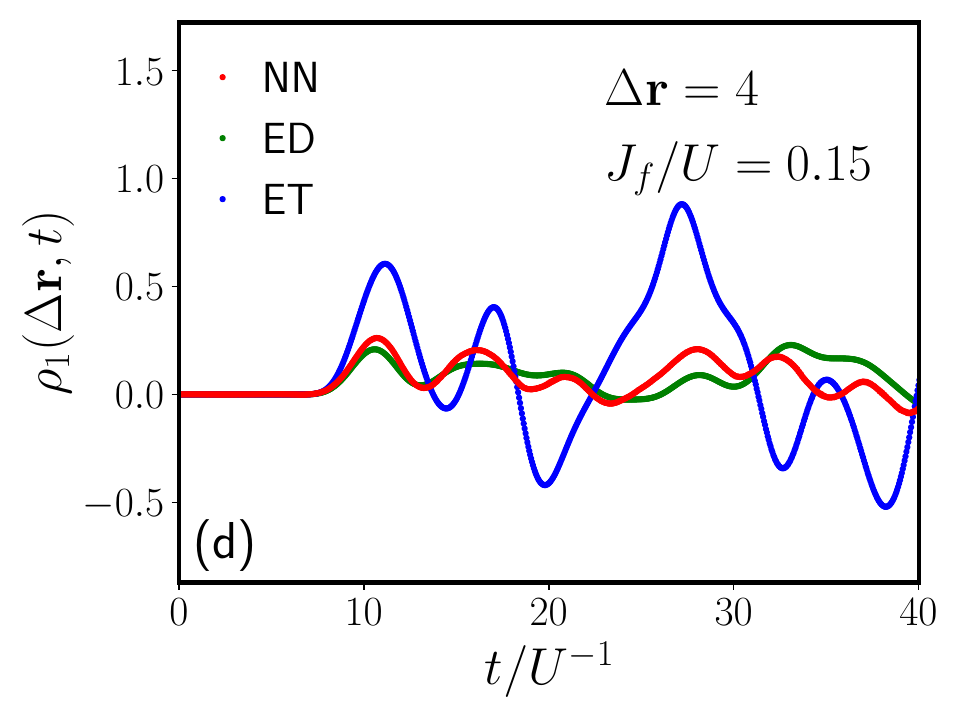}
}
\subfigure{
\includegraphics[width=0.3\textwidth]{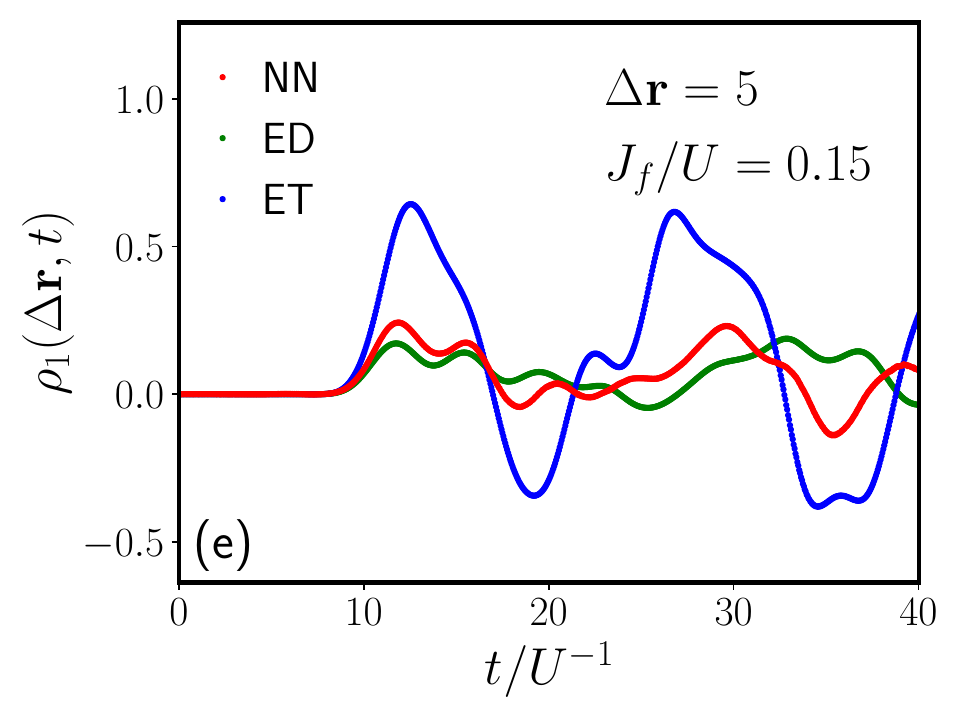}
}
\subfigure{
\includegraphics[width=0.3\textwidth]{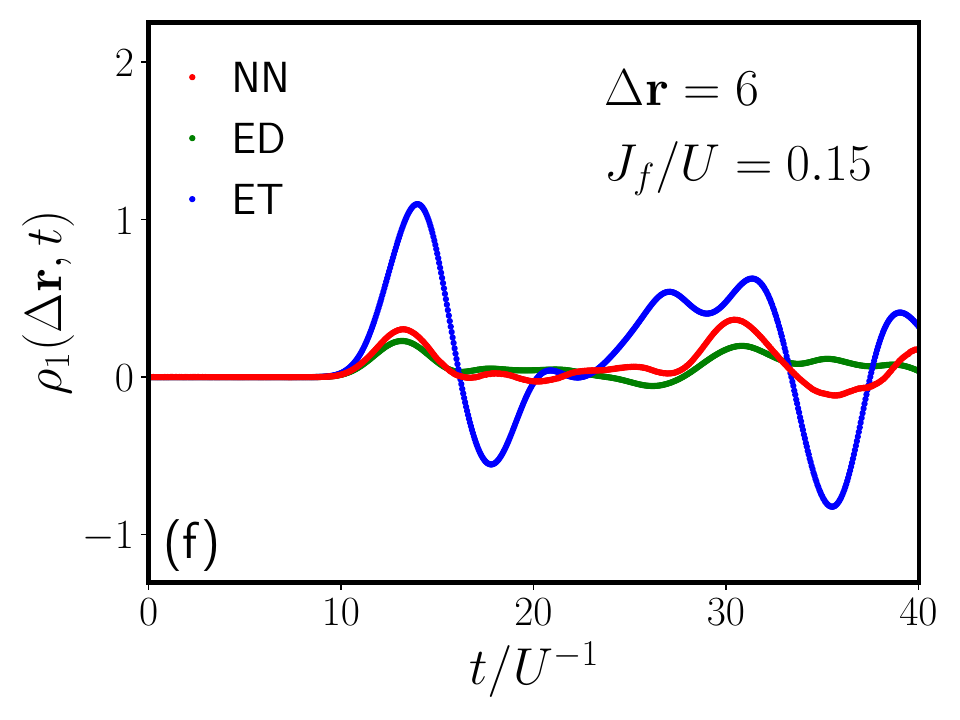}
}
\caption[Comparing NN, ED, and ET for J=0.07 and L=13]{Comparison of the single-particle density matrix $\rho_1\left(\Delta\mathbf{r}, t\right)$ obtained from the effective theory (ET), exact diagonalization (ED) and the neural network (NN) model for a one-dimensional lattice with length 13. Sub-figures (a) to (f) correspond to displacement vectors $\Delta\mathbf{r}$ with magnitudes of 1 to 6, respectively. The remaining parameters used were $\mu/U = 0.42$, $J_f/U = 0.15$, $t_c /U^{-1} = 5$, and $t_Q/ U^{-1} = 0.1$. }
\label{fig: NN 13 J=0.15}
\end{figure}

\end{widetext}

\subsection{Results in one dimension} 
Figure \ref{fig: comparing ED ET NN interapolation} presents a comparison between the ET, ED, and the NN predictions for $L=10$, $\rho_1(\Delta \textbf{r}=1, t)$, and $J_f/U$ values of 0.02, 0.05, and 0.08. For extrapolation testing, we generated new data with the ET for $\rho_1(\Delta \textbf{r}, t)$ at $J_f/U=0.07$ and $L = 14$. We choose $L=14$ specifically because, while it is computationally feasible for ED, it falls outside the range of our training data. 
Figure \ref{fig: NN 14 J=0.07} presents a comparison between the ED, ET, and the NN predictions for $L=14$, $J_f/U=0.07$, and $\Delta \mathbf{r}$ values ranging from 1 to 6. The NN results show significant improvements with respect to the ET results, especially for the first peak and first trough. 

Next, we compare another set of data outside the training data range. We choose $L=13$ but this time $J_f/U=0.15$. This point is near the tip of the Mott lobe and very close to the Mott insulator-superfluid phase boundary. The ET result is less accurate than for smaller $J_f/U$ in predicting the amplitude of $\rho_1\left(\Delta\mathbf{r}, t\right)$. Figure \ref{fig: NN 13 J=0.15} presents a comparison between the ED, ET, and the NN predictions for $L=13$, $J_f/U=0.15$, and $\Delta \mathbf{r}$ values ranging from 1 to 6. Again the NN results show an improvement upon the ET, most notably for the prediction of the amplitude of the first peak. 

The decrease in model accuracy at higher hopping amplitudes is due to the greater deviation of the ET from the ED at higher \(J_f\). Consequently, to improve the NN's accuracy in predicting ED results at higher hopping amplitudes, more data points are required, especially for these higher values. Note that our choice of hopping amplitudes is uniformly distributed, which results in more accurate predictions at lower hopping amplitudes. Therefore, increasing the dataset size can enhance the model's accuracy. In Appendix \ref{app:effect_reducing_dataset_size}, we demonstrate how increasing the number of data points by raising the number of \(J_f\) points reduces the training loss.

We next applied our NN to a regime beyond the reach of ED calculations. Our focus is on a one-dimensional lattice with a length of 20, and we set $J_f/U=0.04$. In such a setting with small hopping amplitudes, our expectation from Fig. \ref{fig: ED ET comparing} is that the ET results should be very close to those from ED. Specifically, we expect the ED calculation of $\rho_1\left(\Delta\mathbf{r}, t\right)$ to exhibit a slightly smaller amplitude, by about $5\%$ at the first peak. Figure \ref{fig: NN 20 J=0.04} presents a comparison between the ET and NN predictions for $L=20$, $J_f/U=0.04$, and $\Delta \mathbf{r}$ values ranging from 1 to 3. There is a slight reduction in the amplitude of the first peak, especially for the displacement vector $\Delta\mathbf{r}=1$.

Next, we explored the case where $L=20$ and $J_f/U=0.07$. From Fig. \ref{fig: ED ET comparing}, we expect a more significant adjustment in the amplitude of $\rho_1\left(\Delta\mathbf{r}, t\right)$ for these
parameters than in Fig.~\ref{fig: NN 20 J=0.04}. Figure \ref{fig: NN 20 J=0.07} provides a comparison between the ET and NN predictions for these parameters, specifically for $L=20$, $J_f/U=0.07$, and $\Delta \mathbf{r}$ values from 1 to 3. There is a larger adjustment to amplitude of the first peak of $\rho_1\left(\Delta\mathbf{r}, t\right)$ in Fig. \ref{fig: NN 20 J=0.07} as compared to Fig. \ref{fig: NN 20 J=0.04}.

The U-Net architecture allows our NN model to typically outperform the ET approximation. 
Indications from Fig.~\ref{fig: NN 20 J=0.07} suggest that this is also true outside the training region.

\begin{widetext}

\begin{figure}[thb]
\subfigure{
\includegraphics[width=0.3\textwidth]{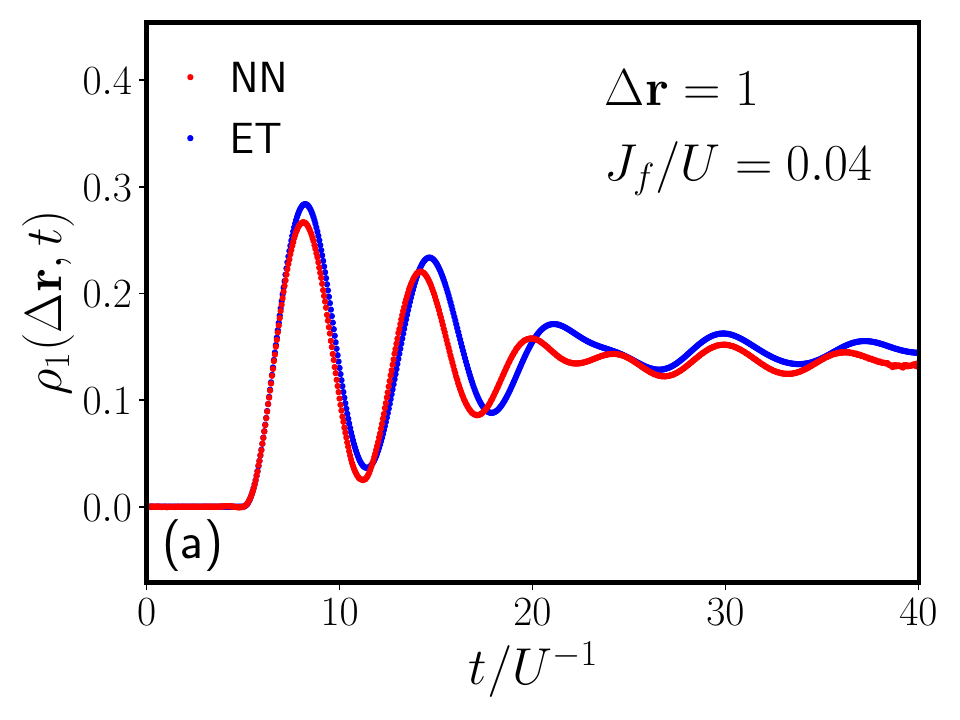}
}
\subfigure{
\includegraphics[width=0.3\textwidth]{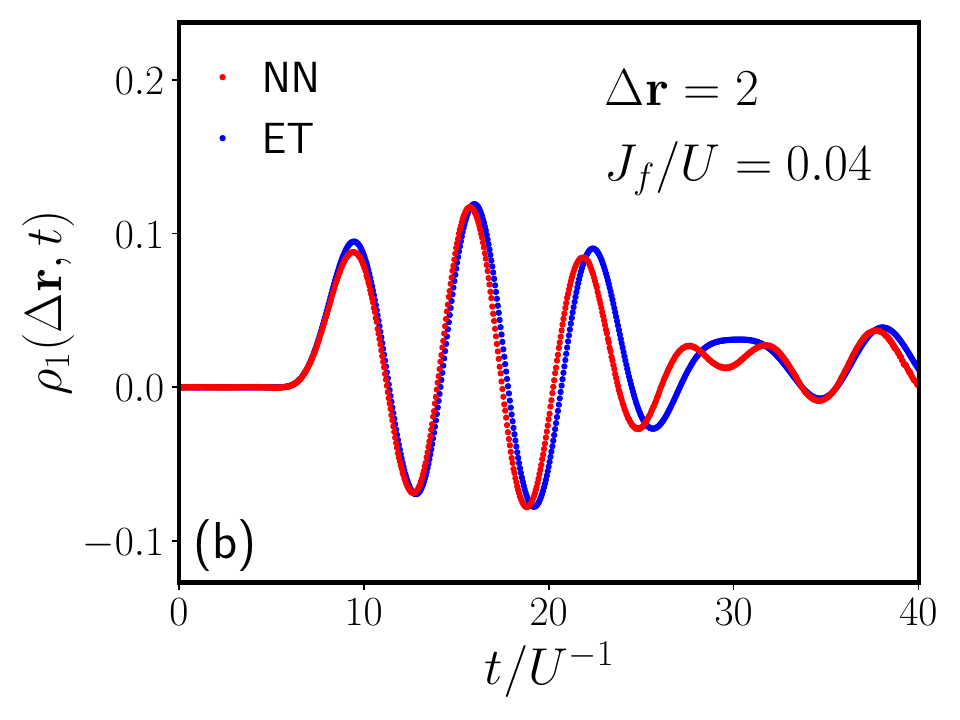}
}
\subfigure{
\includegraphics[width=0.3\textwidth]{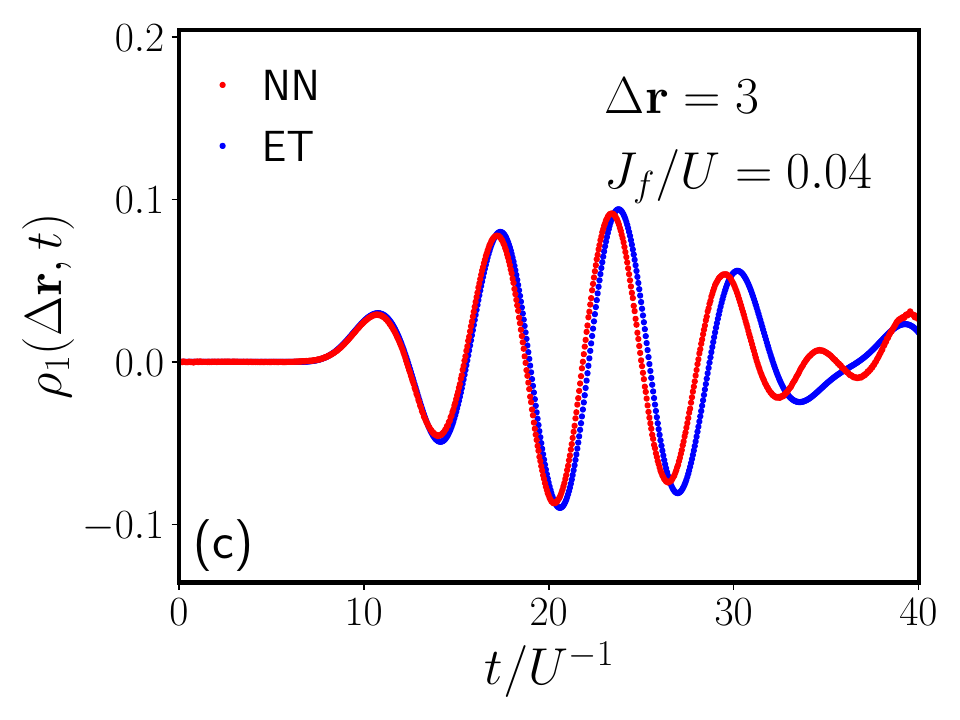}
}
\caption{Comparison of the single-particle density matrix $\rho_1\left(\Delta\mathbf{r}, t\right)$ obtained from the effective theory (ET) and the neural network (NN) model for a one-dimensional lattice with length 20. Sub-figures (a) to (c) correspond to displacement vectors $\Delta\mathbf{r}$ with magnitudes of 1 to 3, respectively. The remaining parameters used were $\mu/U = 0.42$, $J_f/U = 0.04$, $t_c /U^{-1} = 5$, and $t_Q/ U^{-1} = 0.1$. }
\label{fig: NN 20 J=0.04}
\end{figure}

\begin{figure}[htb]
\subfigure{
\includegraphics[width=0.3\textwidth]{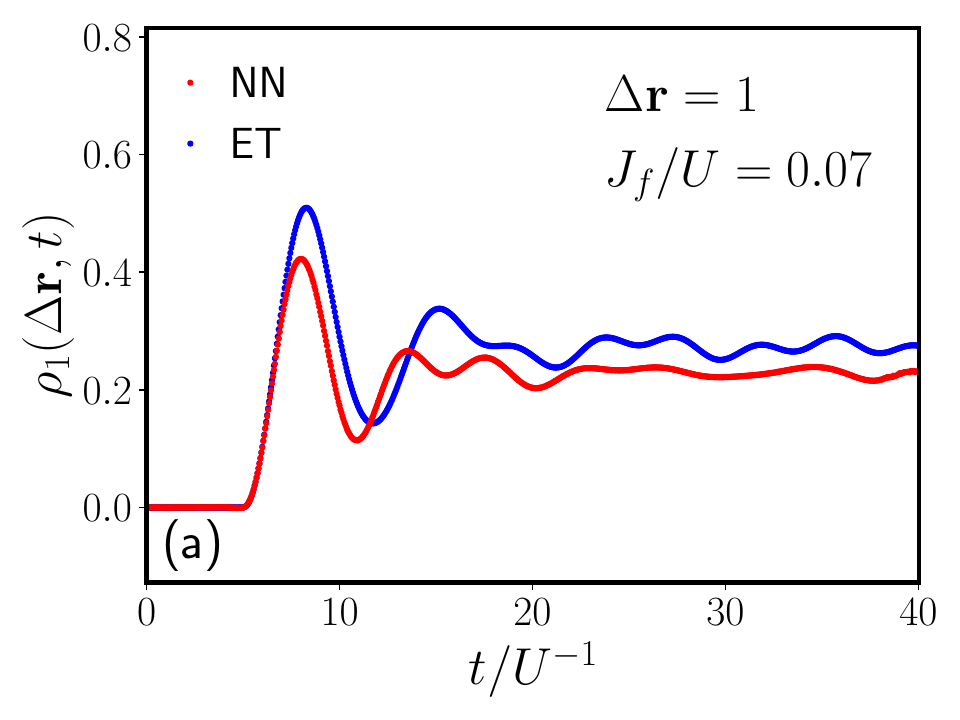}
}
\subfigure{
\includegraphics[width=0.3\textwidth]{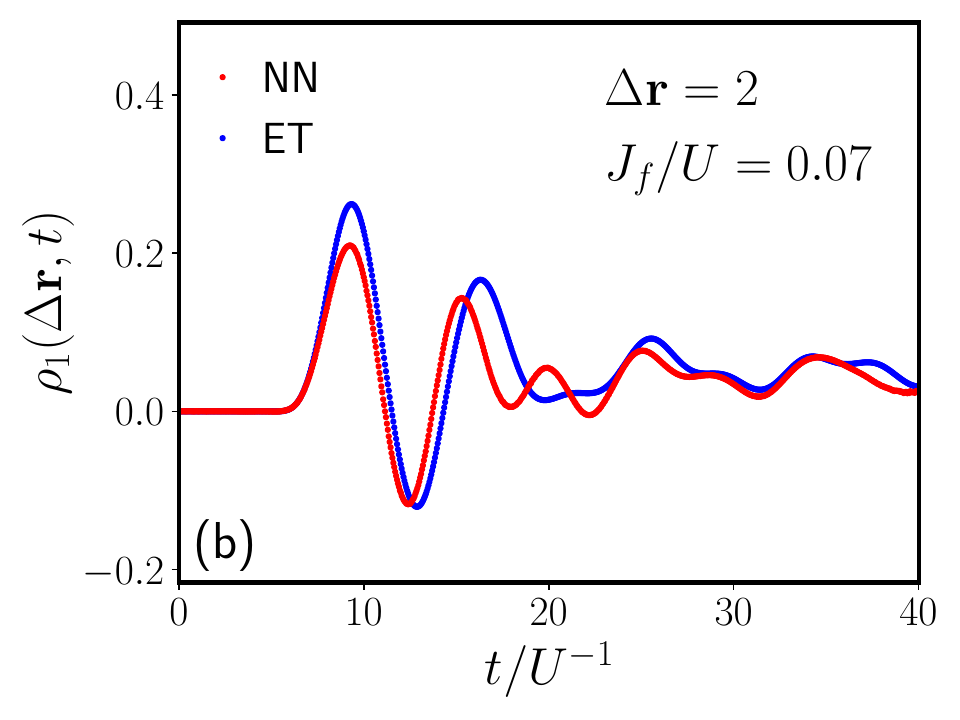}
}
\subfigure{
\includegraphics[width=0.3\textwidth]{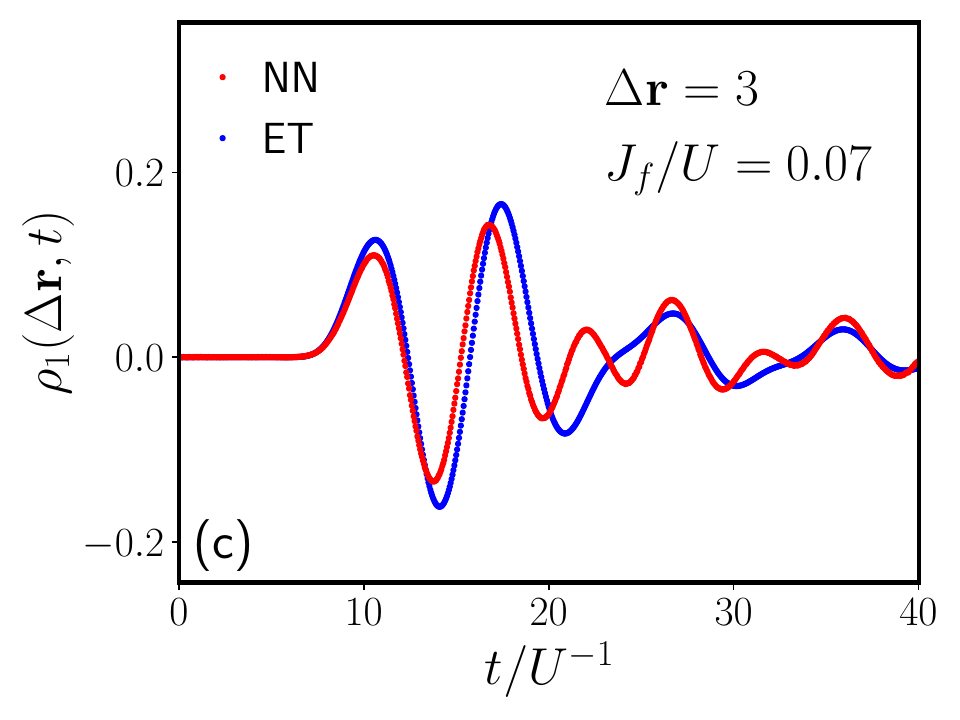}
}
\caption{Comparison of the single-particle density matrix $\rho_1\left(\Delta\mathbf{r}, t\right)$ obtained from the effective theory (ET) and the neural network (NN) model for a one-dimensional lattice with length 20. Sub-figures (a) to (c) correspond to displacement vectors $\Delta\mathbf{r}$ with magnitudes of 1 to 3, respectively. The remaining parameters used were $\mu/U = 0.42$, $J_f/U = 0.07$, $t_c /U^{-1} = 5$, and $t_Q/ U^{-1} = 0.1$. }
\label{fig: NN 20 J=0.07}
\end{figure}

\end{widetext}

\begin{widetext}

\begin{figure}[htb]
\subfigure{
\includegraphics[width=0.3\textwidth]{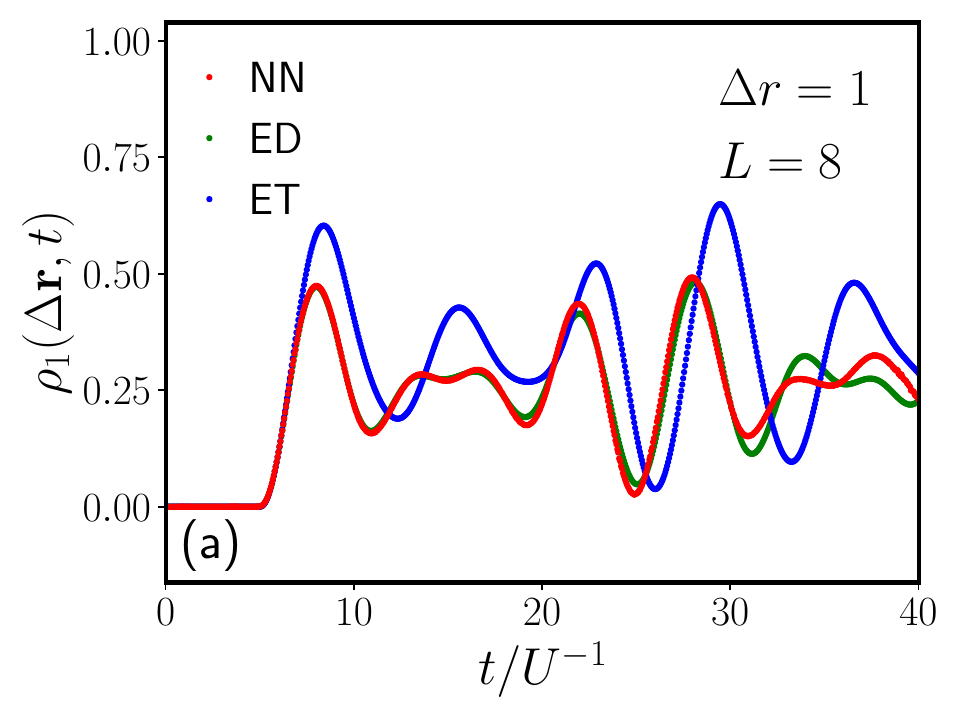}
}
\subfigure{
\includegraphics[width=0.3\textwidth]{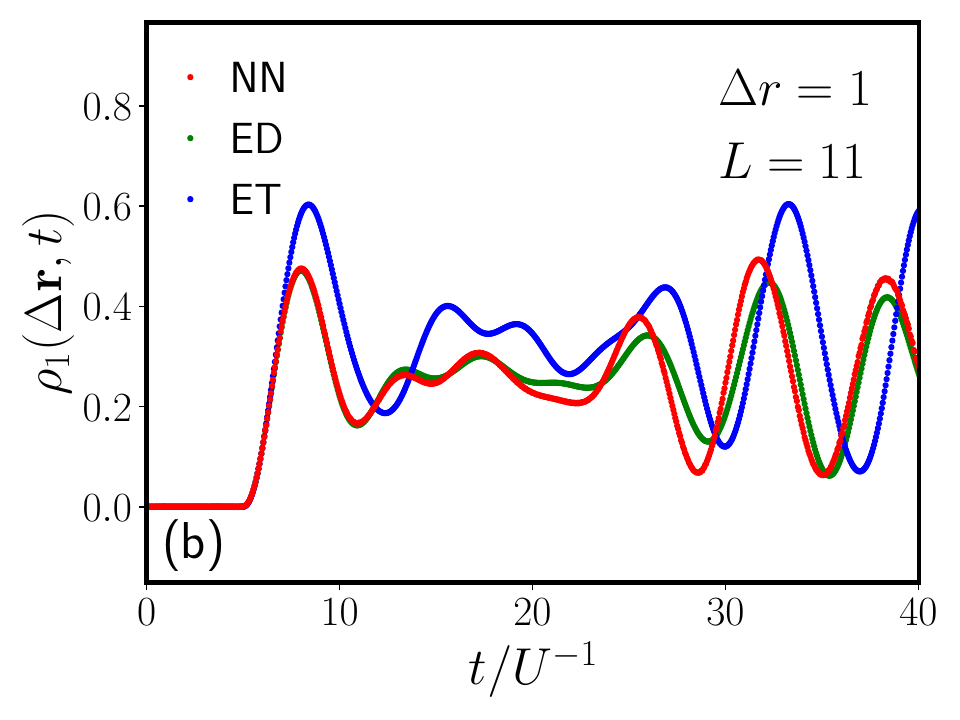}
}
\subfigure{
\includegraphics[width=0.3\textwidth]{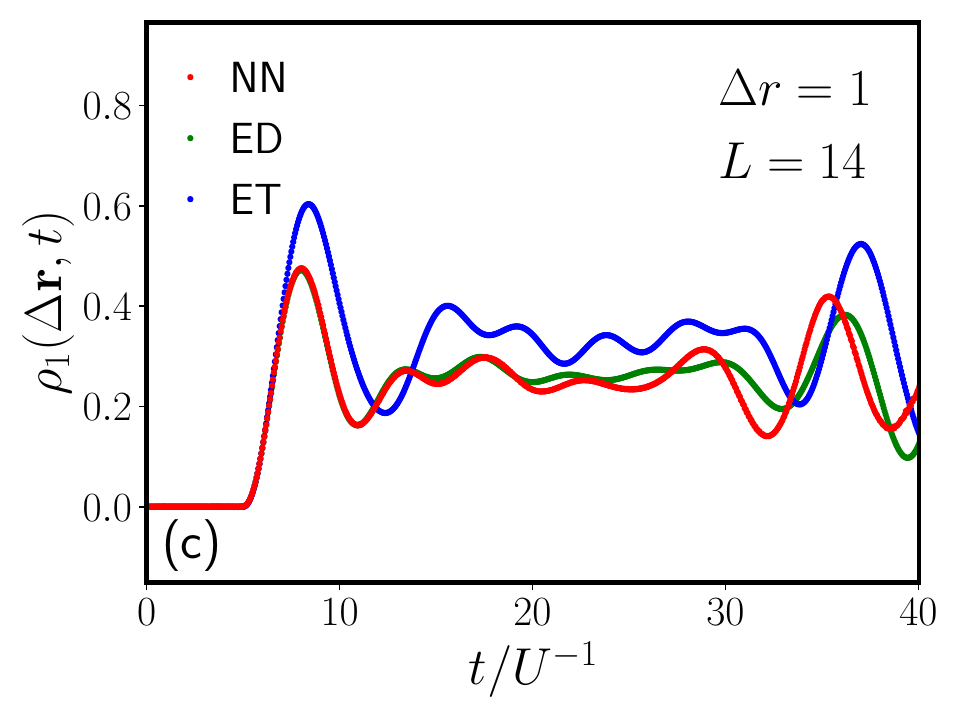}
}
\caption{Comparison of the single-particle density matrix $\rho_1\left(\Delta\mathbf{r}, t\right)$ obtained from the effective theory (ET) and neural network (NN) model (trained
for $L \leq 7$), for a one-dimensional lattice the displacement vector $\Delta\mathbf{r}=1$. Sub-figures correspond to lattice lengths of  $L=8$, $L=11$, and $L=14$, respectively. The parameters used were $J_f/U=0.07$, $\mu/U = 0.42$, $t_c /U^{-1} = 5$, and $t_Q/ U^{-1} = 0.1$. }
\label{fig: train over L from 2 to 7}
\end{figure}
\end{widetext}

Finally, to evaluate the model's ability to extrapolate to larger system sizes, we trained the same model using 24,000 data points from \(L=2\) to \(L=7\)
and then used this model to predict $\rho_1$ for $J_f = 0.07$ with $L = 8, 11$ and $14$, as illustrated in Fig. \ref{fig: train over L from 2 to 7}. 
The results indicate that the model can extrapolate effctively to at least twice the system size. It is important to note that, as the dataset size reduces, as explained in Appendix \ref{app:effect_reducing_dataset_size}, the accuracy of the model decreases compared to when it was trained on 60,000 data points.

\noindent 

\begin{widetext}

\begin{figure}[htb]
\subfigure{
\includegraphics[width=0.3\textwidth]{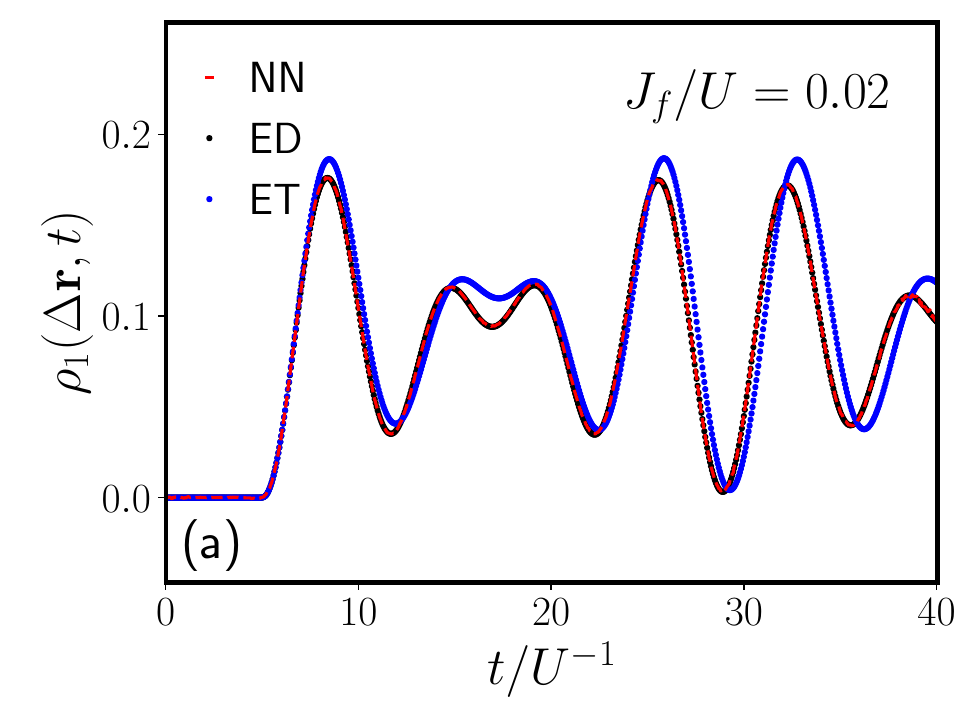}
}
\subfigure{
\includegraphics[width=0.3\textwidth]{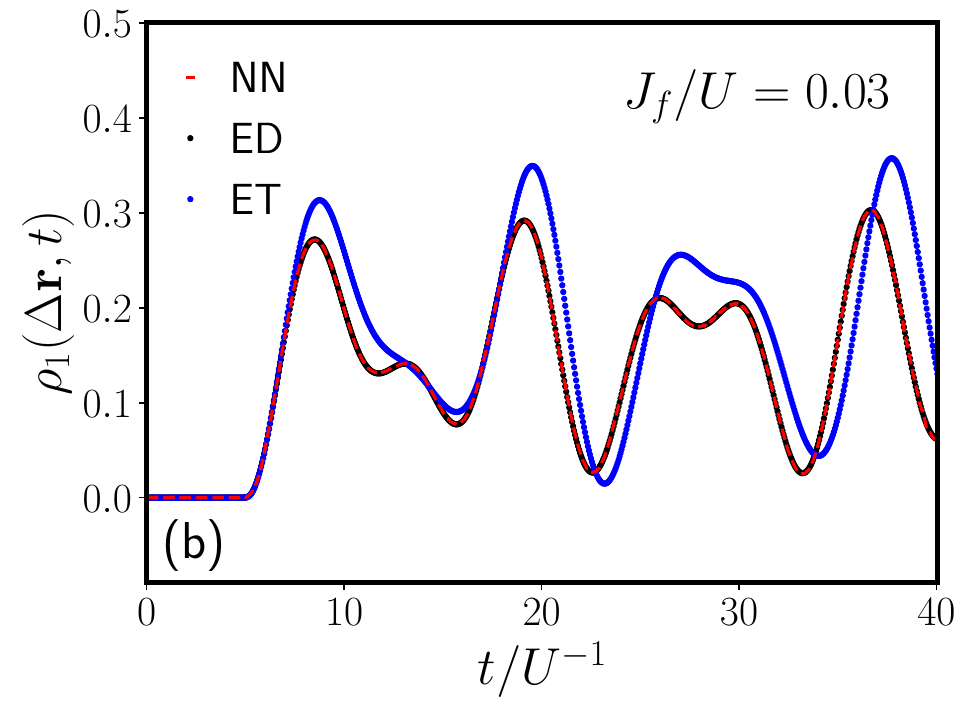} 
}
\subfigure{
\includegraphics[width=0.3\textwidth]{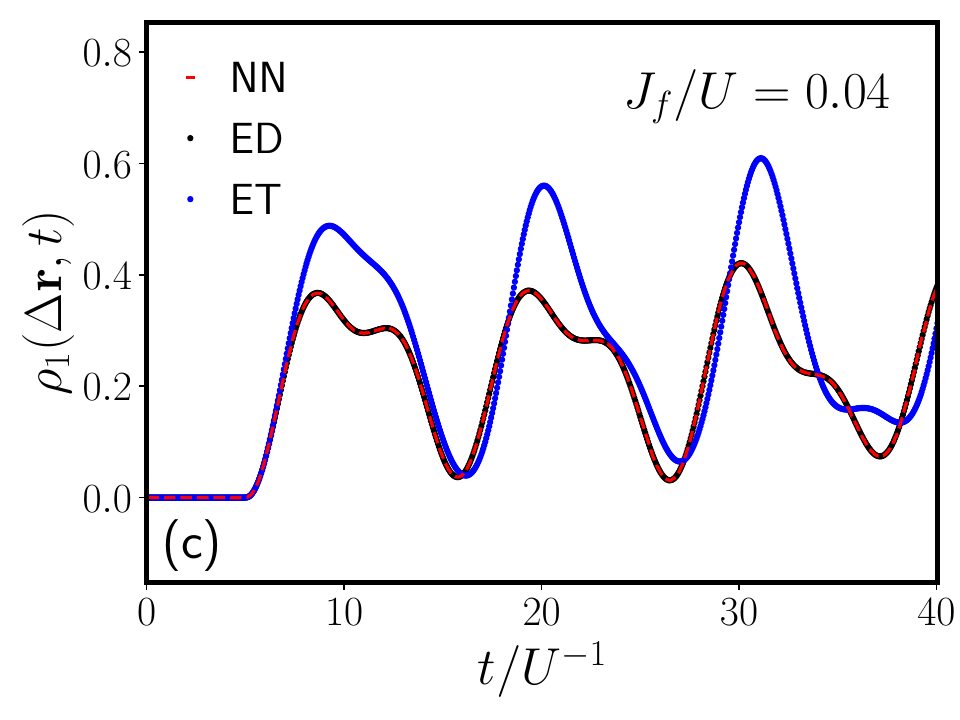}
}
\caption{Comparison of the single-particle density matrix $\rho_1\left(\Delta\mathbf{r}, t\right)$ obtained from the effective theory (ET), exact diagonalization (ED) and the neural network (NN) model for a $3 \times 3$ lattice with $\Delta\mathbf{r}=(1,0)$. Sub-figures (a) to (d) correspond to hopping amplitudes $J_f/U$ with magnitudes of 0.02, 0.04, and 0.05, respectively. The remaining parameters used were $\mu/U = 0.42$, $t_c /U^{-1} = 5$, and $t_Q/ U^{-1} = 0.1$. }
\label{fig: comparing ED ET NN interapolation 2D}
\end{figure}

\begin{figure}[htb]
\subfigure{
\includegraphics[width=0.3\textwidth]{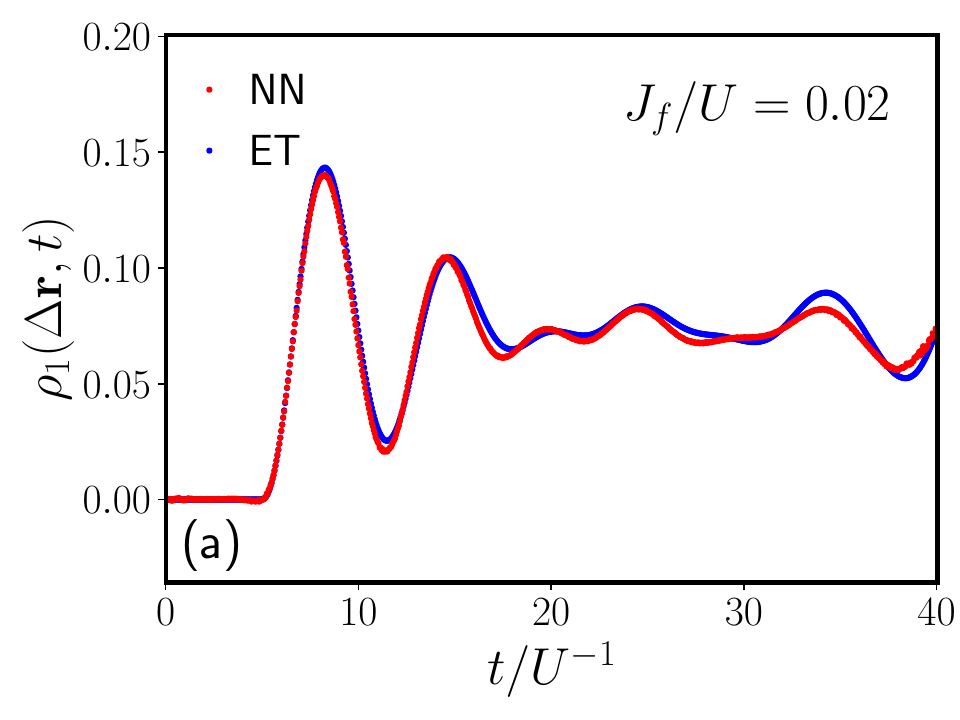}
}
\subfigure{
\includegraphics[width=0.3\textwidth]{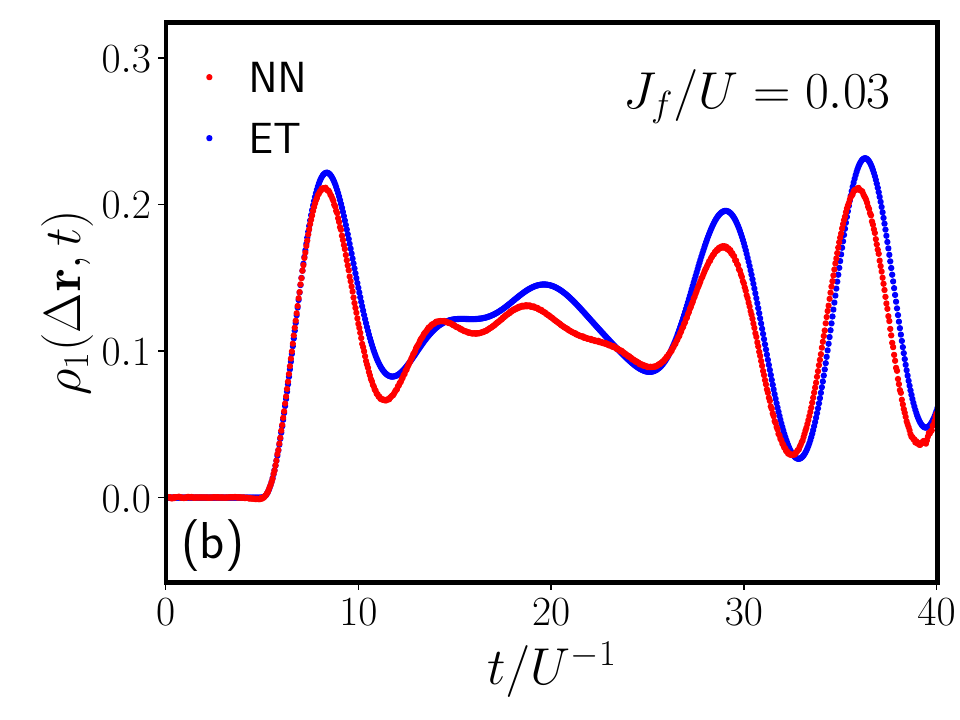}
}
\subfigure{
\includegraphics[width=0.3\textwidth]{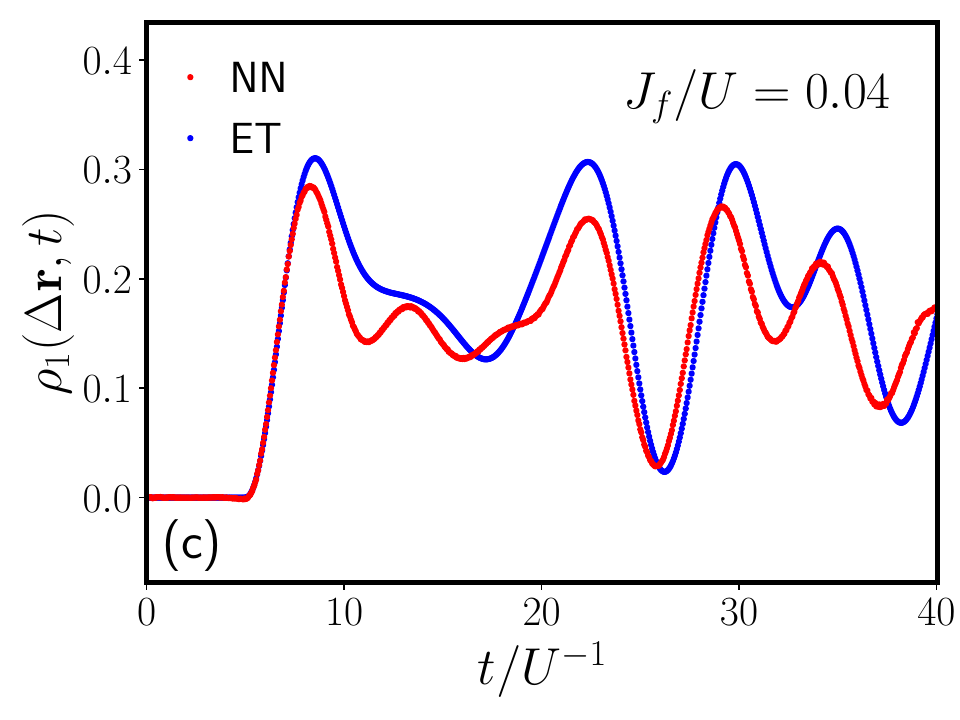}
}
\caption{Comparison of the single-particle density matrix $\rho_1\left(\Delta\mathbf{r}, t\right)$ obtained from the effective theory (ET) and neural network (NN) model for a two-dimensional $4 \times 4$ lattice with the displacement vector $\Delta\mathbf{r}=(1,0)$. Sub-figures (a) to (c) correspond to hopping amplitudes of  $J_f/U=0.02$, $0.03$, and $0.04$, respectively. The parameters used were $\mu/U = 0.42$, $t_c /U^{-1} = 5$, and $t_Q/ U^{-1} = 0.1$. }
\label{fig: NN 2d 4}
\end{figure}

\begin{figure}[htb]
\subfigure{
\includegraphics[width=0.3\textwidth]{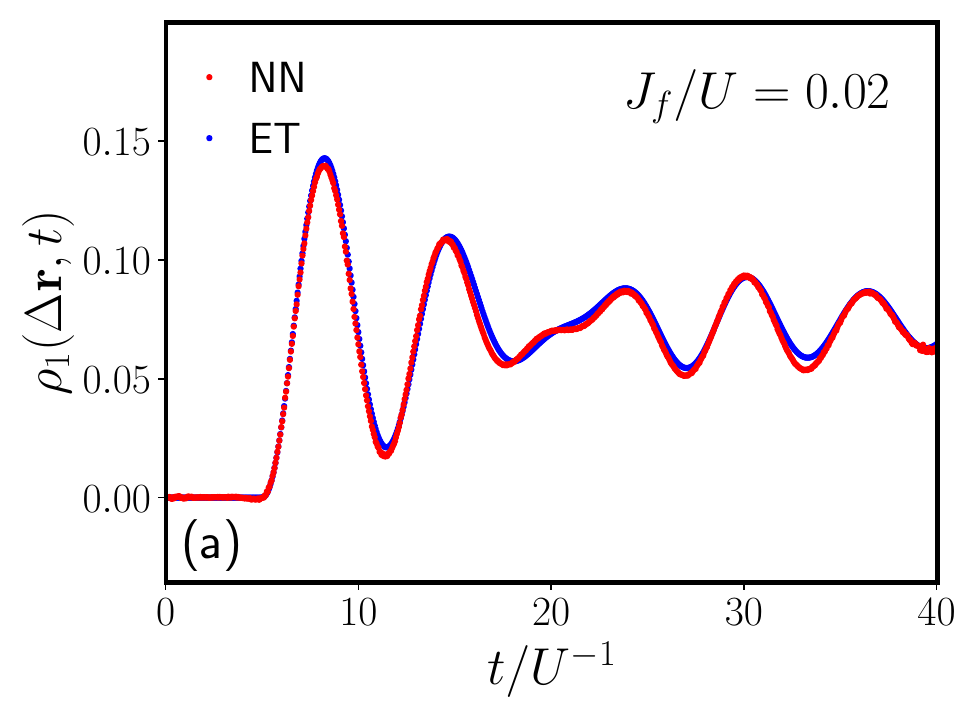}
}
\subfigure{
\includegraphics[width=0.3\textwidth]{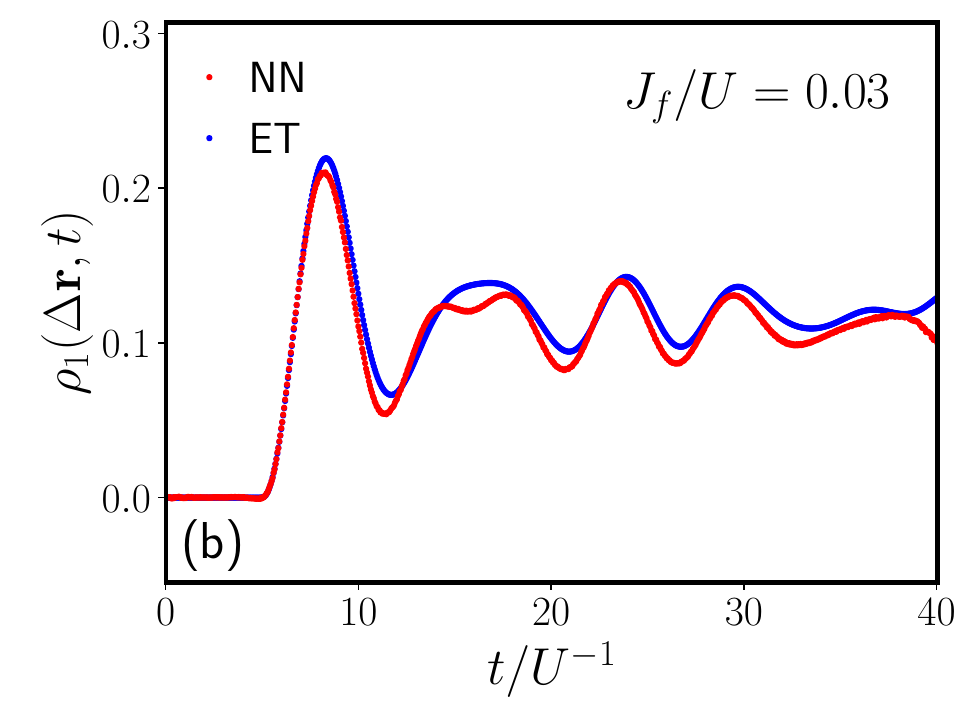}
}
\subfigure{
\includegraphics[width=0.3\textwidth]{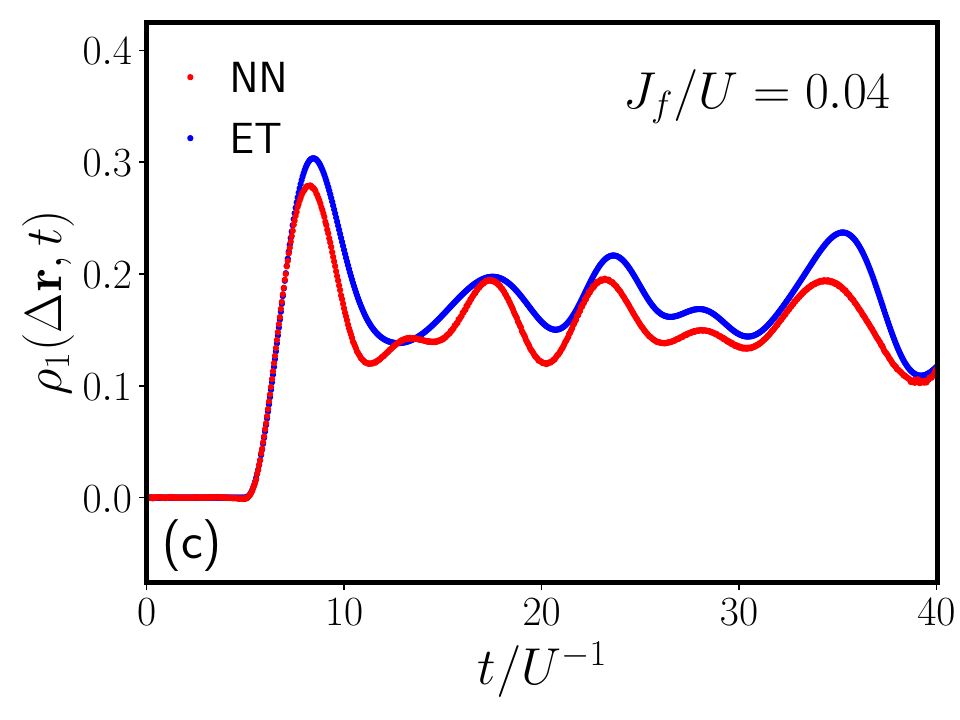}
}
\caption{Comparison of the single-particle density matrix $\rho_1\left(\Delta\mathbf{r}, t\right)$ obtained from the effective theory (ET) and the neural network (NN) model for a two-dimensional $6 \times 6$ lattice with the displacement vector $\Delta\mathbf{r}=(1,0)$. Sub-figures (a) to (c) correspond to hopping amplitudes of  $J_f/U=0.02$, $0.03$, and $0.04$, respectively. The remaining parameters used were $\mu/U = 0.42$, $t_c /U^{-1} = 5$, and $t_Q/ U^{-1} = 0.1$. }
\label{fig: NN 2d 6}
\end{figure}

\end{widetext}

\subsection{Two Dimensions}
For two dimensions, we used a similar training approach to the one we used in one dimension.  We generated ET and ED data for square lattices of sizes \(2 \times 2\) and \(3 \times 3\). We sampled the 
parameter \(J_f/U\) uniformly between \(0.0001\) and \(0.052\), yielding \(10,384\) unique data samples. Figure \ref{fig: comparing ED ET NN interapolation 2D} presents a comparison between the ED, ET, and the NN predictions for $\rho_1(\Delta\bvec{r},t)$ for a \(3 \times 3\) lattice with $\Delta \textbf{r} =\left(1,0\right)$, and $J_f/U$ values of 0.02, 0.04, and 0.05 showing that the NN is able to reproduce the ED data
for given ET input in the training region.

Here we utilize the same U-Net architecture with a similar learning rate, number of training cycles (epochs), batch size, test set to validation set ratio, and optimization method as in the one-dimensional case.

Similarly to the one-dimensional case, in two-dimensions, we anticipate that the NN will output a smaller amplitude for the 
single-particle density matrix $\rho_1\left(\Delta\mathbf{r}, t\right)$ as compared to the ET as the hopping amplitude increases. 
Figure \ref{fig: NN 2d 4} shows a comparison between the ET and NN output for the parameters $L=4$, $\Delta\mathbf{r}= \left(1, 0\right)$, and hopping amplitudes of $J_f/U=0.02$, $0.03$, and $0.04$, exhibiting similar behaviour to one dimension.

We also considered a larger two-dimensional lattice to observe the effect of increascing the $L$ on the NN results. Figure \ref{fig: NN 2d 6} displays a comparison between the ET and NN results for lattices with $L=6$, $\Delta\mathbf{r}= (1, 0)$, and hopping amplitudes $J_f/U=0.02$, $0.03$, and $0.04$.

\begin{widetext}

\begin{figure}[htb]
\subfigure{
\includegraphics[width=0.3\textwidth]{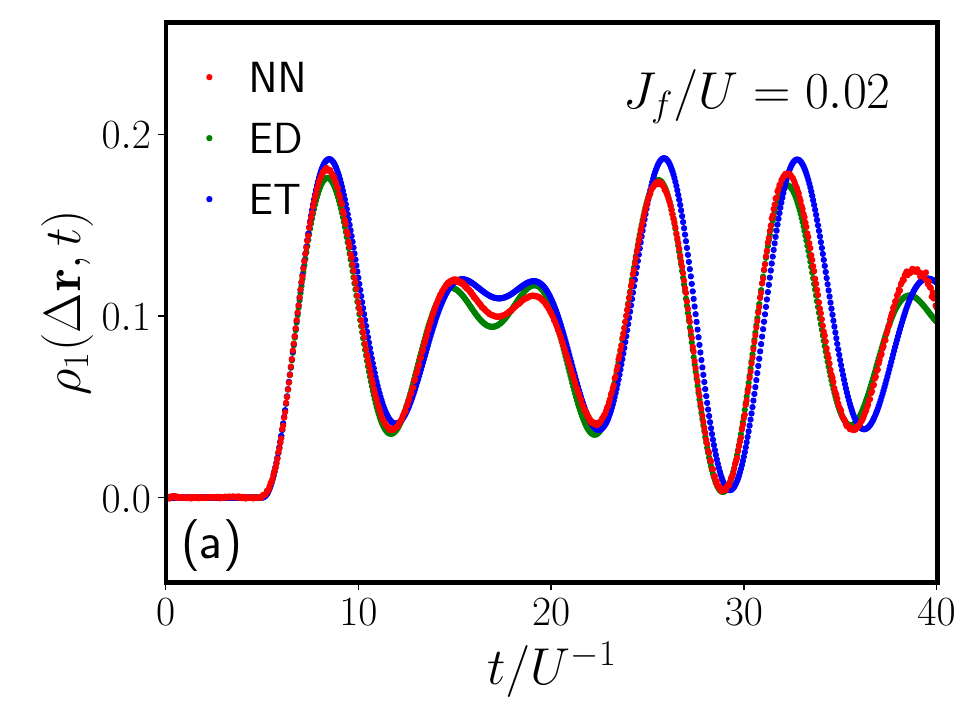}
}
\subfigure{
\includegraphics[width=0.3\textwidth]{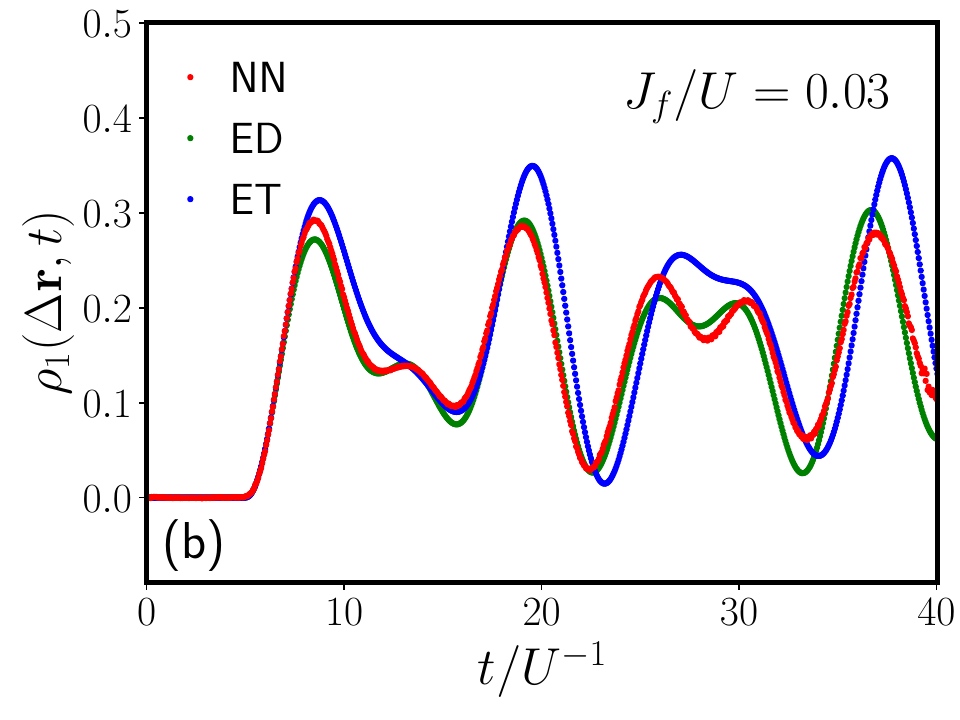}
}
\subfigure{
\includegraphics[width=0.3\textwidth]{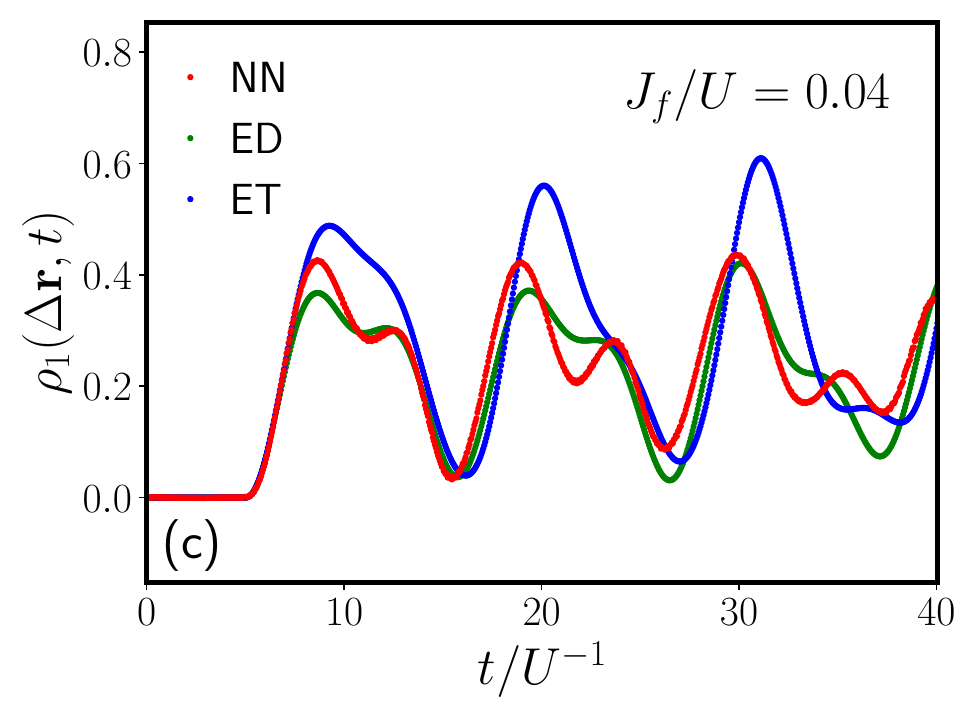}
}
\caption{Comparison of the single-particle density matrix $\rho_1\left(\Delta\mathbf{r}, t\right)$ obtained from the effective theory (ET) and the neural network (NN) model (trained on a $2\times 2$ lattice) for a two-dimensional $3 \times 3$ lattice with the displacement vector $\Delta\mathbf{r}=(1,0)$. Sub-figures (a) to (c) correspond to hopping amplitudes of  $J_f/U=0.02$, $0.03$, and $0.04$, respectively. The remaining parameters used were $\mu/U = 0.42$, $t_c /U^{-1} = 5$, and $t_Q/ U^{-1} = 0.1$.}
\label{fig: train over 2 by 2 lattice}
\end{figure}
\end{widetext}

Finally, similarly to the one-dimensional case, we evaluated the model's ability to extrapolate to larger system sizes. We trained the same model with similar training parameters, but this time using 5,192 data points from $L=2$ to determine whether it can extrapolate to $L=3$. Figure \ref{fig: train over 2 by 2 lattice} presents a comparison between the ED, ET, and NN predictions for $\rho_1(\Delta \mathbf{r}, t)$ for a $3 \times 3$ lattice with $\Delta \mathbf{r} = (1,0)$, and $J_f/U$ values of 0.02, 0.04, and 0.05. Although the model was trained on data from only one lattice size $2 \times 2$ and with a relatively small dataset, it still manages to predict some of the complex features fairly accurately.

% We observed that as the lattice size increases, the result of the NN model approaches that of the ET. 
% %As noted above, one of the features of the U-net training is that if it is unable to improve upon the input, it retains that input as output.
% **** WILL REWRITE THIS IN LIGHT OF NEW DATA ****** 
% We find that for system sizes larger than about twice the maximum system size in the training data, the NN returns similar results to the ET. This suggests that the NN is effective up to system sizes close to twice those considered in the training but likely does not improve on the ET for larger system sizes.
% ***************

\section{Discussion and Conclusions}
\label{sec:disc}
In this paper, we explored using neural network-enhanced analysis to calculate out-of-equilibrium single-particle density correlations for the Bose-Hubbard model.
The main result in our work is that we are able to use the combination of the results of our approximate effective theory and a U-Net neural network to obtain 
essentially exact results for the single particle density matrix, both inside and outside the training region in parameter space.  

For one dimension our training 
data covered up to $L=11$, we found excellent agreement with ED data for $L=14$ and we suggest that our method likely improves on the ET up to at least $L=20$ for weak hopping.
In two dimensions, the accessible range of parameters was more limited than in one dimension. Similarly to one dimension we observed that the NN results generally show 
a reasonable improvement over the ET for lattice lengths up to twice the maximum lattice length in our training data.  
Although we were limited in system sizes where we could apply ED, the sizes investigated are comparable to those studied in experiments \cite{Cheneau2012,Takasu2020}. 

We observed that as the hopping amplitude increased, the ET deviated more from the ED results, making it harder for the NN model to accurately predict the correct behaviour. To resolve this, we found that more data points are required, especially for higher $J_f$ values, so that the model can better learn how to adjust the input to obtain the correct output. This also applies to lattice displacement vectors, as the ET results deviated more from the ED results for some values compared to others. Again, we argue that this can be improved by having a larger dataset.

Interestingly, we also observed that the model can extrapolate the corrections even with access to data points from very small system sizes. In one dimension, by having access to data points up to $L=7$, it could extrapolate at least up to $L=14$. In two dimensions, just by being trained on data points from a $2 \times 2$ lattice, it could predict the $3 \times 3$ results. Furthermore, one can always improve the final results by training the model on a larger dataset.

Increasing the dataset size by considering more $J_f$ values exhibits a linear time dependency with respect to its size. This is generally not a limiting factor in many-body problems. In contrast, increasing the dataset size by considering more values of $L$, has an exponential time dependency, particularly for the ED method. This distinction implies that one can feasibly expand the dataset in $J_f$ but not $L$ in a reasonable amount of time to achieve better results.

It is unclear to what extent the accuracy of the effective theory limits the performance of the NN-enhanced approach.  The original motivation for this work was that the 
amplitude of the single particle density matrix calculated for larger values of $J_f/U$ is not quantitatively accurate.  However, the output of the U-Net NN at these values of 
$J_f/U$ is dramatically improved for system sizes larger than those where the NN was trained ($L=13$ versus $L=11$), as illustrated in Figs.~\ref{fig: NN 13 J=0.15} and
\ref{fig: train over L from 2 to 7}.  The extent to which dimensionality affects the efficacy of the NN model is still an open question.
Certainly our results in two dimensions appear to be equally promising as those in 
one dimension, albeit at smaller linear dimensions. 

It appears that the ingredients that allow the application of a U-Net NN architecture to be effective for out-of-equilibrium correlations in the BHM are: 
i) the ability to obtain exact results (using ED) for small system sizes, and ii) an approximate effective model that gives qualitatively correct behaviour 
of the dynamics for larger system sizes.  This approach should be just as useful for obtaining out-of-equilibrium dynamics in other quantum many body systems.
Given that exact diagonalization is generally readily available for small systems, e.g. by using the QuSpin library \cite{QuSpin}, for a wide
variety of time-dependent and time-independent Hamiltonians, the main barrier to using the approach we have outlined appears to be the existence of a 
reasonably accurate effective theory.

An example of a system that satisfies the criteria laid out above is the transverse field Ising model.  This can be diagonalized exactly for relatively small system sizes (up to about 6$\times$6 in two dimensions).  Effective descriptions, including mean field theory, matrix product states, and tensor network approaches also exist.  Training a U-Net 
in the way we have here might be helpful for improving benchmarking of results obtained from quantum hardware, e.g. as in Ref.~\cite{king2024}.

\vspace{0.5 in}

\section{Acknowledgements}
A. M-J. and M. P. K. were supported by NSERC.  We thank Jean-Sebastien Bernier for helpful discussions and Matthew Fitzpatrick for insightful feedback on the manuscript 
and helpful suggestions.
\vspace{0.5 cm}

\appendix

\section{Comparison of Training and Validation Losses}
\label{app:training_vs_validation_loss}
The dataset was partitioned into training and validation subsets, comprising 95\% and 5\% of the total, respectively. This distribution allocates 57,000 instances to the training set and 3,000 to the validation set. The optimization strategy effectively refined the learning rate to approximately \(1.56 \times 10^{-6}\) by the conclusion of the eighth epoch. As detailed in Table \ref{tab:training_summary}, the progression of the average training loss over mini-batches, the training loss computed over the entire dataset at the end of each epoch, and the validation loss over the course of eight epochs is documented.

\begin{table}[ht]
\centering
\setlength{\tabcolsep}{4pt}
\begin{tabular}{llll}
\toprule
\textbf{LR} & \textbf{Avg Loss} & \textbf{Val Loss} & \textbf{End Train Loss}\\
\midrule
$1.00\times 10^{-4}$ & $8.38\times 10^{-4}$ & $5.93\times 10^{-5}$ & $6.03\times 10^{-5}$ \\
$1.00\times 10^{-4}$ & $8.07\times 10^{-5}$ & $2.81\times 10^{-5}$ & $3.02\times 10^{-5}$ \\
$5.00\times 10^{-5}$ & $9.26\times 10^{-6}$ & $6.91\times 10^{-6}$ & $7.02\times 10^{-6}$ \\
$2.50\times 10^{-5}$ & $2.56\times 10^{-6}$ & $1.43\times 10^{-6}$ & $1.43\times 10^{-6}$ \\
$1.25\times 10^{-5}$ & $1.02\times 10^{-6}$ & $1.84\times 10^{-6}$ & $1.86\times 10^{-6}$ \\
$6.25\times 10^{-6}$ & $6.71\times 10^{-7}$ & $6.08\times 10^{-7}$ & $6.00\times 10^{-7}$ \\
$3.13\times 10^{-6}$ & $5.58\times 10^{-7}$ & $5.39\times 10^{-7}$ & $5.27\times 10^{-7}$ \\
$1.56\times 10^{-6}$ & $5.18\times 10^{-7}$ & $5.19\times 10^{-7}$ & $5.09\times 10^{-7}$ \\
\bottomrule
\end{tabular}
\caption{Summary of learning rates, average training loss over mini-batches, validation loss, and training loss at the end of each epoch}
\label{tab:training_summary}
\end{table}

\begin{figure}
\centering
\includegraphics[width=0.44\textwidth]{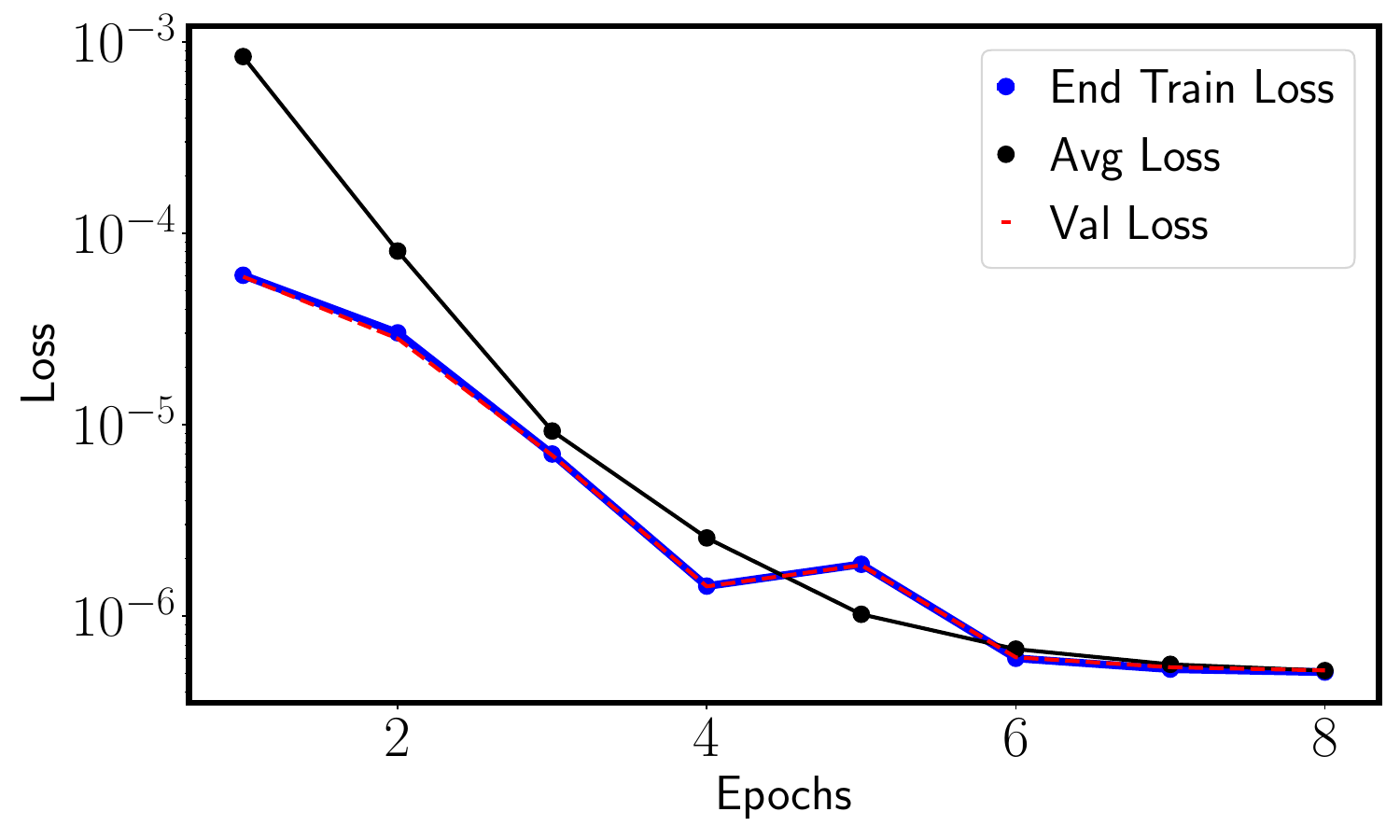}
\caption{Illustration of the average training loss over mini-batches, training loss at the end of each epoch, and validation loss for eight epochs.}
\label{fig:loss_figure}
\end{figure}

Regularization was not required in our model as no overfitting was observed during the initial 8 epochs; both the training and validation set errors converged to approximately \(5 \times 10^{-7}\). Figure \ref{fig:loss_figure} shows the average training loss over mini-batches, the training loss at the end of each epoch, and the validation loss over the 8 epochs in the logarithmic scale.

\section{Effect of Reducing Dataset Size on Loss}
\label{app:effect_reducing_dataset_size}
In this Appendix, we investigated the impact of systematically reducing the dataset size on the model's loss. Starting from an initial dataset size of 60,000 instances, we progressively halved the dataset size down to 30,000, 15,000, 7,500, 3,750 and 1,875. This reduction was achieved by halving the Jf concentrations at each step.

% Regularization was not required in our model as no overfitting was observed during the initial 8 epochs; both the training and validation set errors converged to approximately \(7 \times 10^{-7}\).

\begin{figure}
\subfigure{
\includegraphics[width=0.44\textwidth]{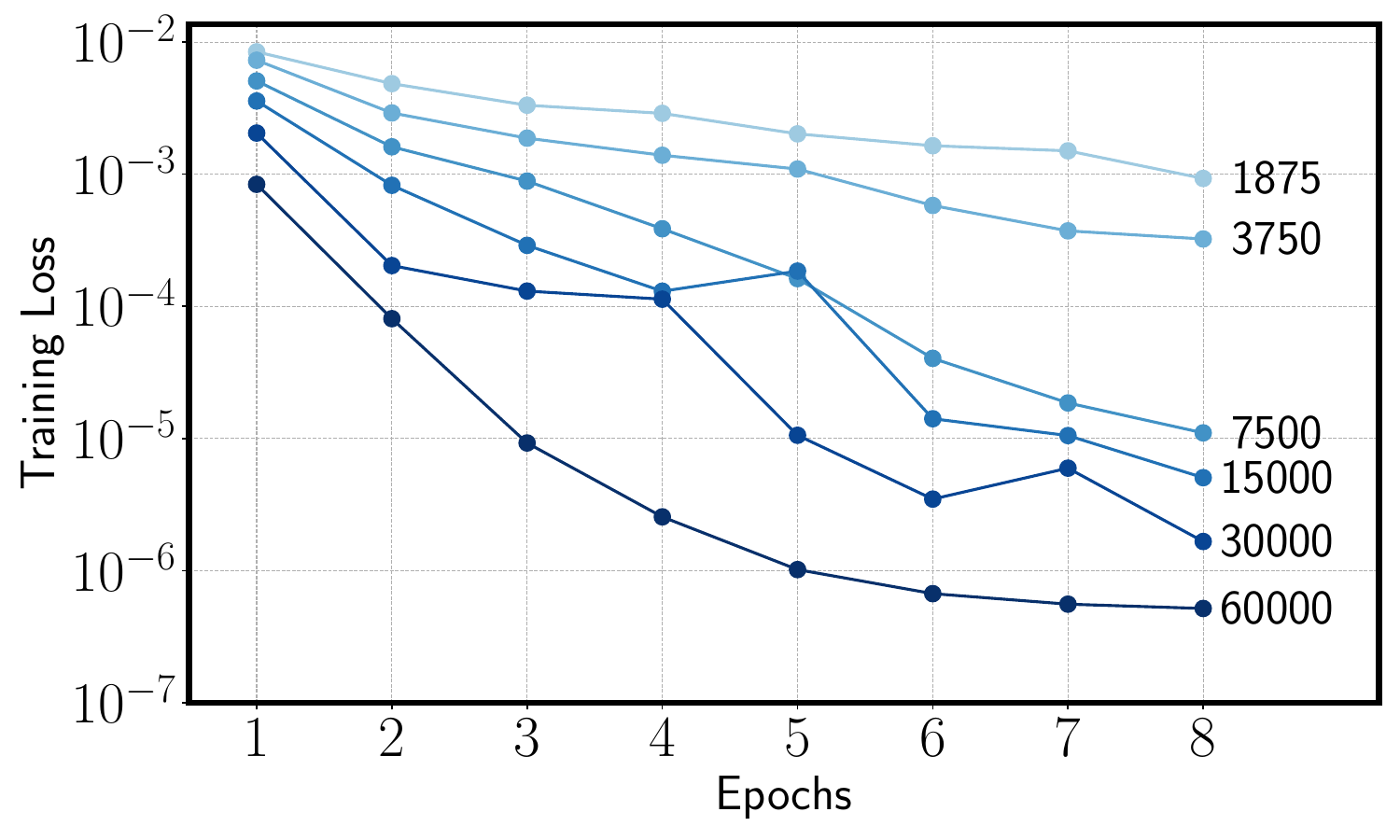}
}
\caption{demonstrating training losses for different dataset sizes over 8 epochs. }
\label{fig:train_loss_reductions}
\end{figure}

\begin{figure}
\subfigure{
\includegraphics[width=0.44\textwidth]{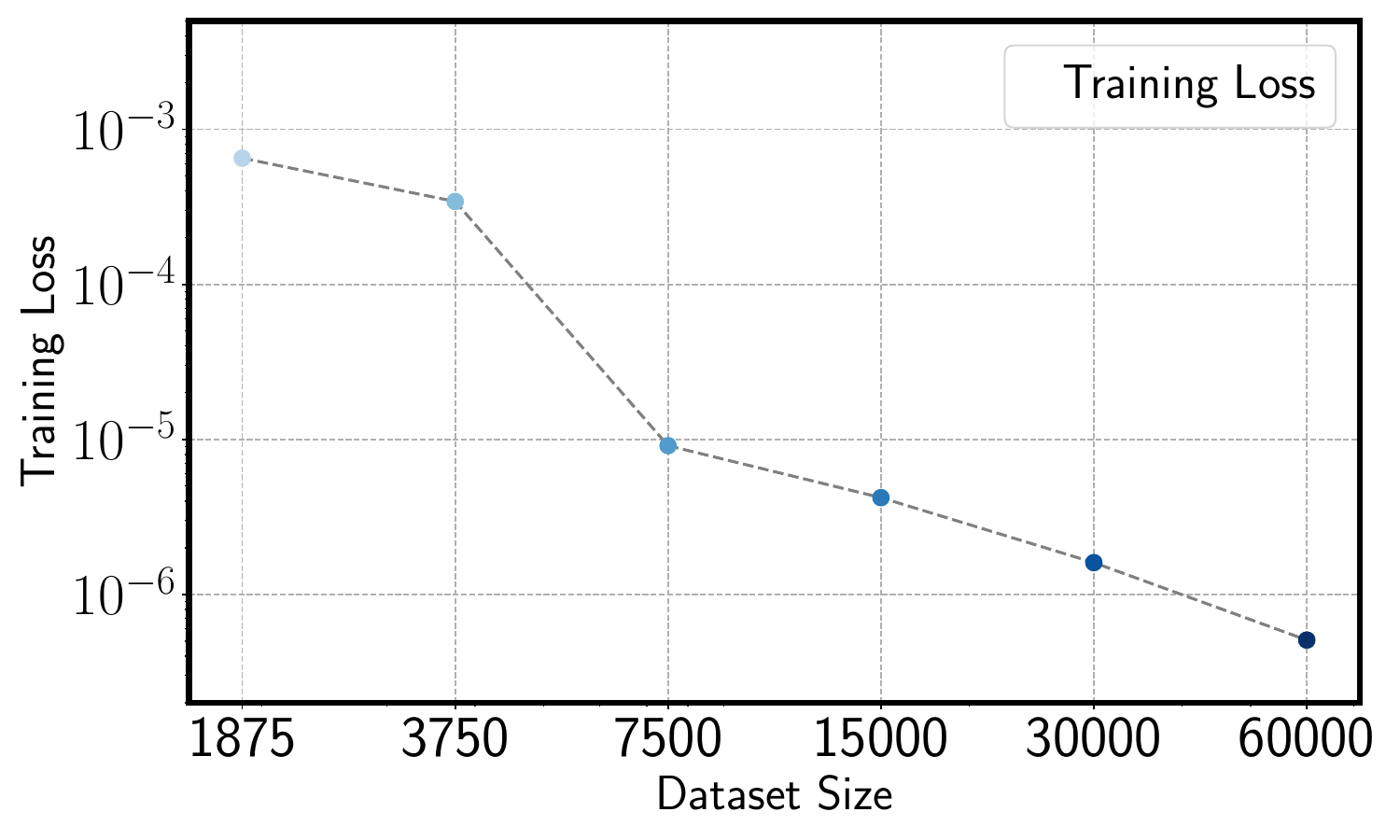}
}

\caption{Final training loss for different dataset sizes over 8 epochs.}
\label{fig:Final_loss_reduction}
\end{figure}

Figure \ref{fig:train_loss_reductions} shows the training loss of the same neural network over 8 epochs for 6 different dataset sizes: 1,875, 3,750, 7,500, 15,000, 30,000, and 60,000. Figure \ref{fig:Final_loss_reduction} shows the final training loss at the 8th epoch for these 6 datasets, demonstrating how the training loss decreases as the dataset size increases.

% \begin{widetext}

% \begin{figure}
% \subfigure{
% \includegraphics[width=0.44\textwidth]{loss_comparizon.pdf}
% }
% \subfigure{
% \includegraphics[width=0.44\textwidth]{loss_decreasing_by_data_size.pdf}
% }

% \caption{Left: Training loss for different dataset sizes over 8 epochs, and right: final training loss for different dataset sizes over 8 epochs. Dataset sizes are 1875, 3750, 7500, 15000, 30000, and 60000.}
% \label{fig:train_loss_reduction}
% \end{figure}

% \end{widetext}

The results indicate that the loss values generally increase as the dataset size decreases, suggesting that a smaller dataset size can reduce the model's ability to learn effectively.

% \section{Results for $L=13$}
% \label{app:L_13_results}
% In this appendix, we present the extrapolation results for \(L=13\). Figure \ref{fig: NN_13_J=0.07} presents a comparison between the ED, ET, and the NN predictions for \(L=13\), \(J_f/U=0.07\), and \(\Delta \mathbf{r}\) values ranging from 1 to 6. The NN results show significant improvements with respect to the ET results.
% \begin{widetext}
    
% \begin{figure}
% \centering
% \subfigure{
% \includegraphics[width=0.3\textwidth]{Comparing_L=13_Jf=0.07_r=1.pdf}
% }
% \subfigure{
% \includegraphics[width=0.3\textwidth]{Comparing_L=13_Jf=0.07_r=2.pdf}
% }
% \subfigure{
% \includegraphics[width=0.3\textwidth]{Comparing_L=13_Jf=0.07_r=3.pdf}
% }
% \subfigure{
% \includegraphics[width=0.3\textwidth]{Comparing_L=13_Jf=0.07_r=4.pdf}
% }
% \subfigure{
% \includegraphics[width=0.3\textwidth]{Comparing_L=13_Jf=0.07_r=5.pdf}
% }
% \subfigure{
% \includegraphics[width=0.3\textwidth]{Comparing_L=13_Jf=0.07_r=6.pdf}
% }
% \caption{Comparison of the single-particle density matrix $\rho_1\left(\Delta\mathbf{r}, t\right)$ obtained from the effective theory (ET), exact diagonalization (ED) and the neural network (NN) model for a one-dimensional lattice with length 13. Sub-figures (a) to (f) correspond to displacement vectors $\Delta\mathbf{r}$ with magnitudes of 1 to 6, respectively. The remaining parameters used were $\mu/U = 0.42$, $J_f/U = 0.07$, $t_c /U^{-1} = 5$, and $t_Q/ U^{-1} = 0.1$. }
% \label{fig: NN_13_J=0.07}
% \end{figure}

% \end{widetext}

\bibliographystyle{apsrev4-1}
\bibliography{arxiv}

\end{document}